\documentstyle[12pt,floatfig,epsfig,citequo]{article}
\epsfverbosetrue
\newcommand{\beq}{\begin{equation}}
\newcommand{\eeq}{\end{equation}}
\newcommand{\beqn}{\begin{eqnarray}}
\newcommand{\eeqn}{\end{eqnarray}}
\newcommand{\beqns}{\begin{eqnarray*}}
\newcommand{\eeqns}{\end{eqnarray*}}
\newcommand{\vs}{\\[0.3cm]\indent}

\newcommand{\hm}{\hspace{-0.05cm}}

\newcommand{\intl}{\int\limits}

\def\NP{Nucl. Phys.}
\def\PL{Phys. Lett.}
\def\PR{Phys. Rev.}

\def\PRL{Phys. Rev. Lett.}

\def\ZP{Z. Phys.}

\def\ea{{\it et al.}}
\def\Cl{Collaboration}
\def\sfs{spectral functions}

\def\asm{$\alpha_s(M_\tau^2)$}

\def\ms{$m_s$}
\def\msm{$m_s(M_\tau^2)$}

\def\mss{$m_s(s)$}

\def\ee{$e^+e^-$}

\def\RtS{$R_{\tau,S}$}

\def\GG{$\langle(\alpha_s/\pi) GG\rangle$}

\def\MSbar{$\overline{\rm MS}$}
\def\FOPTCI{$\rm FOPT_{\rm CI}$}

\begin{document}

\begin{center}
{\large EUROPEAN LABORATORY FOR PARTICLE PHYSICS}
\end{center}

\vspace{0.5cm}
\begin{flushright}CERN-EP/99-026\\
22 February 1999\\
\end{flushright}

\begin{center}
\vspace{0.5cm}
{\huge
\bf
Study of $\bf{\tau}$ decays involving kaons, spectral 
functions and determination of the strange quark mass}
\end{center}

\begin{center}
\vspace{0.5cm}
\bf {The ALEPH Collaboration}
\end{center}

\vspace{0.5cm}
\begin{abstract}
All ALEPH measurements of branching ratios of $\tau$ decays involving kaons 
are summarized including a combination of results obtained with 
$K^0_S$ and $K^0_L$ detection. The decay dynamics are studied, leading to
the determination of contributions from vector 
$K^*(892)$ and $K^{*}(1410)$, 
and axial-vector $K_1(1270)$ and $K_1(1400)$ resonances.
Agreement with isospin symmetry is observed among the different
final states. Under the hypothesis of the conserved vector current,
the spectral function for the $K\overline{K}\pi$ mode is compared 
with the corresponding cross section for low energy $e^+e^-$ 
annihilation, yielding an axial-vector fraction of $(94^{\,+6}_{\,-8})\%$ 
for this mode. The branching ratio for $\tau$ decay into all strange final
states is determined to be 
$B(\tau^-\to X^-(S=-1)\nu_\tau)=(28.7\pm1.2)\times 10^{-3}$. 
The measured mass spectra of the strange $\tau$ decay modes are exploited 
to derive the $S=-1$ spectral function. A combination of strange and 
nonstrange spectral functions is used to determine the strange 
quark mass and nonperturbative contributions to the strange hadronic width.
A method is developed to avoid the bad convergence of the
spin zero hadronic component, with the result 
$m_s(M_\tau^2)=(176^{\,+46}_{\,-57})$ MeV/$c^2$. The evolution down to
1~GeV gives $m_s(1~{\rm GeV}^2) = 
(234^{\,+61}_{\,-76})~{\rm MeV}/c^2$.
\end{abstract}
\vspace{1cm}
\centerline{\it (To be submitted to European Physical Journal C) }
\pagestyle{empty}
\newpage
\small
%
\newlength{\saveparskip}
\newlength{\savetextheight}
\newlength{\savetopmargin}
\newlength{\savetextwidth}
\newlength{\saveoddsidemargin}
\newlength{\savetopsep}
\setlength{\saveparskip}{\parskip}
\setlength{\savetextheight}{\textheight}
\setlength{\savetopmargin}{\topmargin}
\setlength{\savetextwidth}{\textwidth}
\setlength{\saveoddsidemargin}{\oddsidemargin}
\setlength{\savetopsep}{\topsep}
%
%
\setlength{\parskip}{0.0cm}
\setlength{\textheight}{25.0cm}
\setlength{\topmargin}{-1.5cm}
\setlength{\textwidth}{16 cm}
\setlength{\oddsidemargin}{-0.0cm}
\setlength{\topsep}{1mm}
\pretolerance=10000
\centerline{\large\bf The ALEPH Collaboration}
\footnotesize
\vspace{0.5cm}
{\raggedbottom
\begin{sloppypar}
\samepage\noindent
R.~Barate,
D.~Decamp,
P.~Ghez,
C.~Goy,
\mbox{J.-P.~Lees},
E.~Merle,
\mbox{M.-N.~Minard},
B.~Pietrzyk
\nopagebreak
\begin{center}
\parbox{15.5cm}{\sl\samepage
Laboratoire de Physique des Particules (LAPP), IN$^{2}$P$^{3}$-CNRS,
F-74019 Annecy-le-Vieux Cedex, France}
\end{center}\end{sloppypar}
\vspace{2mm}
\begin{sloppypar}
\noindent
R.~Alemany,
M.P.~Casado,
M.~Chmeissani,
J.M.~Crespo,
E.~Fernandez,
\mbox{M.~Fernandez-Bosman},
Ll.~Garrido,$^{15}$
E.~Graug\`{e}s,
A.~Juste,
M.~Martinez,
G.~Merino,
R.~Miquel,
Ll.M.~Mir,
A.~Pacheco,
I.C.~Park,
I.~Riu
\nopagebreak
\begin{center}
\parbox{15.5cm}{\sl\samepage
Institut de F\'{i}sica d'Altes Energies, Universitat Aut\`{o}noma
de Barcelona, E-08193 Bellaterra (Barcelona), Spain$^{7}$}
\end{center}\end{sloppypar}
\vspace{2mm}
\begin{sloppypar}
\noindent
A.~Colaleo,
D.~Creanza,
M.~de~Palma,
G.~Gelao,
G.~Iaselli,
G.~Maggi,
M.~Maggi,
S.~Nuzzo,
A.~Ranieri,
G.~Raso,
F.~Ruggieri,
G.~Selvaggi,
L.~Silvestris,
P.~Tempesta,
A.~Tricomi,$^{3}$
G.~Zito
\nopagebreak
\begin{center}
\parbox{15.5cm}{\sl\samepage
Dipartimento di Fisica, INFN Sezione di Bari, I-70126
Bari, Italy}
\end{center}\end{sloppypar}
\vspace{2mm}
\begin{sloppypar}
\noindent
X.~Huang,
J.~Lin,
Q. Ouyang,
T.~Wang,
Y.~Xie,
R.~Xu,
S.~Xue,
J.~Zhang,
L.~Zhang,
W.~Zhao
\nopagebreak
\begin{center}
\parbox{15.5cm}{\sl\samepage
Institute of High-Energy Physics, Academia Sinica, Beijing, The People's
Republic of China$^{8}$}
\end{center}\end{sloppypar}
\vspace{2mm}
\begin{sloppypar}
\noindent
D.~Abbaneo,
U.~Becker,$^{19}$
G.~Boix,$^{6}$
M.~Cattaneo,
V.~Ciulli,
G.~Dissertori,
H.~Drevermann,
R.W.~Forty,
M.~Frank,
A.W. Halley,
J.B.~Hansen,
J.~Harvey,
P.~Janot,
B.~Jost,
I.~Lehraus,
O.~Leroy,
P.~Mato,
A.~Minten,
A.~Moutoussi,
F.~Ranjard,
L.~Rolandi,
D.~Rousseau,
D.~Schlatter,
M.~Schmitt,$^{20}$
O.~Schneider,$^{23}$
W.~Tejessy,
F.~Teubert,
I.R.~Tomalin,
E.~Tournefier,
A.E.~Wright
\nopagebreak
\begin{center}
\parbox{15.5cm}{\sl\samepage
European Laboratory for Particle Physics (CERN), CH-1211 Geneva 23,
Switzerland}
\end{center}\end{sloppypar}
\vspace{2mm}
\begin{sloppypar}
\noindent
Z.~Ajaltouni,
F.~Badaud,
G.~Chazelle,
O.~Deschamps,
A.~Falvard,
C.~Ferdi,
P.~Gay,
C.~Guicheney,
P.~Henrard,
J.~Jousset,
B.~Michel,
S.~Monteil,
\mbox{J-C.~Montret},
D.~Pallin,
P.~Perret,
F.~Podlyski
\nopagebreak
\begin{center}
\parbox{15.5cm}{\sl\samepage
Laboratoire de Physique Corpusculaire, Universit\'e Blaise Pascal,
IN$^{2}$P$^{3}$-CNRS, Clermont-Ferrand, F-63177 Aubi\`{e}re, France}
\end{center}\end{sloppypar}
\vspace{2mm}
\begin{sloppypar}
\noindent
J.D.~Hansen,
J.R.~Hansen,
P.H.~Hansen,
B.S.~Nilsson,
B.~Rensch,
A.~W\"a\"an\"anen
\begin{center}
\parbox{15.5cm}{\sl\samepage
Niels Bohr Institute, DK-2100 Copenhagen, Denmark$^{9}$}
\end{center}\end{sloppypar}
\vspace{2mm}
\begin{sloppypar}
\noindent
G.~Daskalakis,
A.~Kyriakis,
C.~Markou,
E.~Simopoulou,
I.~Siotis,
A.~Vayaki
\nopagebreak
\begin{center}
\parbox{15.5cm}{\sl\samepage
Nuclear Research Center Demokritos (NRCD), GR-15310 Attiki, Greece}
\end{center}\end{sloppypar}
\vspace{2mm}
\begin{sloppypar}
\noindent
A.~Blondel,
G.~Bonneaud,
\mbox{J.-C.~Brient},
A.~Roug\'{e},
M.~Rumpf,
M.~Swynghedauw,
M.~Verderi,
H.~Videau
\nopagebreak
\begin{center}
\parbox{15.5cm}{\sl\samepage
Laboratoire de Physique Nucl\'eaire et des Hautes Energies, Ecole
Polytechnique, IN$^{2}$P$^{3}$-CNRS, \mbox{F-91128} Palaiseau Cedex, France}
\end{center}\end{sloppypar}
\vspace{2mm}
\begin{sloppypar}
\noindent
E.~Focardi,
G.~Parrini,
K.~Zachariadou
\nopagebreak
\begin{center}
\parbox{15.5cm}{\sl\samepage
Dipartimento di Fisica, Universit\`a di Firenze, INFN Sezione di Firenze,
I-50125 Firenze, Italy}
\end{center}\end{sloppypar}
\vspace{2mm}
\begin{sloppypar}
\noindent
R.~Cavanaugh,
M.~Corden,
C.~Georgiopoulos
\nopagebreak
\begin{center}
\parbox{15.5cm}{\sl\samepage
Supercomputer Computations Research Institute,
Florida State University,
Tallahassee, FL 32306-4052, USA $^{13,14}$}
\end{center}\end{sloppypar}
\vspace{2mm}
\begin{sloppypar}
\noindent
A.~Antonelli,
G.~Bencivenni,
G.~Bologna,$^{4}$
F.~Bossi,
P.~Campana,
G.~Capon,
F.~Cerutti,
V.~Chiarella,
P.~Laurelli,
G.~Mannocchi,$^{5}$
F.~Murtas,
G.P.~Murtas,
L.~Passalacqua,
\mbox{M.~Pepe-Altarelli}$^{1}$
\nopagebreak
\begin{center}
\parbox{15.5cm}{\sl\samepage
Laboratori Nazionali dell'INFN (LNF-INFN), I-00044 Frascati, Italy}
\end{center}\end{sloppypar}
\vspace{2mm}
\begin{sloppypar}
\noindent
L.~Curtis,
J.G.~Lynch,
P.~Negus,
V.~O'Shea,
C.~Raine,
\mbox{P.~Teixeira-Dias},
A.S.~Thompson
\nopagebreak
\begin{center}
\parbox{15.5cm}{\sl\samepage
Department of Physics and Astronomy, University of Glasgow, Glasgow G12
8QQ,United Kingdom$^{10}$}
\end{center}\end{sloppypar}
\vspace{2mm}
\begin{sloppypar}
\noindent
O.~Buchm\"uller,
S.~Dhamotharan,
C.~Geweniger,
P.~Hanke,
G.~Hansper,
V.~Hepp,
E.E.~Kluge,
A.~Putzer,
J.~Sommer,
K.~Tittel,
S.~Werner,$^{19}$
M.~Wunsch
\nopagebreak
\begin{center}
\parbox{15.5cm}{\sl\samepage
Institut f\"ur Hochenergiephysik, Universit\"at Heidelberg, D-69120
Heidelberg, Germany$^{16}$}
\end{center}\end{sloppypar}
\vspace{2mm}
\begin{sloppypar}
\noindent
R.~Beuselinck,
D.M.~Binnie,
W.~Cameron,
P.J.~Dornan,$^{1}$
M.~Girone,
S.~Goodsir,
E.B.~Martin,
N.~Marinelli,
J.K.~Sedgbeer,
P.~Spagnolo,
E.~Thomson,
M.D.~Williams
\nopagebreak
\begin{center}
\parbox{15.5cm}{\sl\samepage
Department of Physics, Imperial College, London SW7 2BZ,
United Kingdom$^{10}$}
\end{center}\end{sloppypar}
\vspace{2mm}
\begin{sloppypar}
\noindent
V.M.~Ghete,
P.~Girtler,
E.~Kneringer,
D.~Kuhn,
G.~Rudolph
\nopagebreak
\begin{center}
\parbox{15.5cm}{\sl\samepage
Institut f\"ur Experimentalphysik, Universit\"at Innsbruck, A-6020
Innsbruck, Austria$^{18}$}
\end{center}\end{sloppypar}
\vspace{2mm}
\begin{sloppypar}
\noindent
A.P.~Betteridge,
C.K.~Bowdery,
P.G.~Buck,
P.~Colrain,
G.~Crawford,
A.J.~Finch,
F.~Foster,
G.~Hughes,
R.W.L.~Jones,
N.A.~Robertson,
M.I.~Williams
\nopagebreak
\begin{center}
\parbox{15.5cm}{\sl\samepage
Department of Physics, University of Lancaster, Lancaster LA1 4YB,
United Kingdom$^{10}$}
\end{center}\end{sloppypar}
\vspace{2mm}
\begin{sloppypar}
\noindent
I.~Giehl,
C.~Hoffmann,
K.~Jakobs,
K.~Kleinknecht,
G.~Quast,
B.~Renk,
E.~Rohne,
\mbox{H.-G.~Sander},
P.~van~Gemmeren,
H.~Wachsmuth,
C.~Zeitnitz
\nopagebreak
\begin{center}
\parbox{15.5cm}{\sl\samepage
Institut f\"ur Physik, Universit\"at Mainz, D-55099 Mainz, Germany$^{16}$}
\end{center}\end{sloppypar}
\vspace{2mm}
\begin{sloppypar}
\noindent
J.J.~Aubert,
C.~Benchouk,
A.~Bonissent,
J.~Carr,$^{1}$
P.~Coyle,
F.~Etienne,
F.~Motsch,
P.~Payre,
M.~Talby,
M.~Thulasidas
\nopagebreak
\begin{center}
\parbox{15.5cm}{\sl\samepage
Centre de Physique des Particules, Facult\'e des Sciences de Luminy,
IN$^{2}$P$^{3}$-CNRS, F-13288 Marseille, France}
\end{center}\end{sloppypar}
\vspace{2mm}
\begin{sloppypar}
\noindent
M.~Aleppo,
M.~Antonelli,
F.~Ragusa
\nopagebreak
\begin{center}
\parbox{15.5cm}{\sl\samepage
Dipartimento di Fisica, Universit\`a di Milano e INFN Sezione di Milano,
I-20133 Milano, Italy}
\end{center}\end{sloppypar}
\vspace{2mm}
\begin{sloppypar}
\noindent
R.~Berlich,
V.~B\"uscher,
H.~Dietl,
G.~Ganis,
K.~H\"uttmann,
G.~L\"utjens,
C.~Mannert,
W.~M\"anner,
\mbox{H.-G.~Moser},
S.~Schael,
R.~Settles,
H.~Seywerd,
H.~Stenzel,
W.~Wiedenmann,
G.~Wolf
\nopagebreak
\begin{center}
\parbox{15.5cm}{\sl\samepage
Max-Planck-Institut f\"ur Physik, Werner-Heisenberg-Institut,
D-80805 M\"unchen, Germany\footnotemark[16]}
\end{center}\end{sloppypar}
\vspace{2mm}
\begin{sloppypar}
\noindent
P.~Azzurri,
J.~Boucrot,
O.~Callot,
S.~Chen,
A.~Cordier,
M.~Davier,
L.~Duflot,
\mbox{J.-F.~Grivaz},
Ph.~Heusse,
A.~H\"ocker,
A.~Jacholkowska,
D.W.~Kim,$^{12}$
F.~Le~Diberder,
J.~Lefran\c{c}ois,
\mbox{A.-M.~Lutz},
\mbox{M.-H.~Schune},
\mbox{J.-J.~Veillet},
I.~Videau,$^{1}$
D.~Zerwas
\nopagebreak
\begin{center}
\parbox{15.5cm}{\sl\samepage
Laboratoire de l'Acc\'el\'erateur Lin\'eaire, Universit\'e de Paris-Sud,
IN$^{2}$P$^{3}$-CNRS, F-91898 Orsay Cedex, France}
\end{center}\end{sloppypar}
\vspace{2mm}
\begin{sloppypar}
\noindent
G.~Bagliesi,
S.~Bettarini,
T.~Boccali,
C.~Bozzi,$^{24}$
G.~Calderini,
R.~Dell'Orso,
I.~Ferrante,
L.~Fo\`{a},
A.~Giassi,
A.~Gregorio,
F.~Ligabue,
A.~Lusiani,
P.S.~Marrocchesi,
A.~Messineo,
F.~Palla,
G.~Rizzo,
G.~Sanguinetti,
A.~Sciab\`a,
G.~Sguazzoni,
R.~Tenchini,
C.~Vannini,
A.~Venturi,
P.G.~Verdini
\samepage
\begin{center}
\parbox{15.5cm}{\sl\samepage
Dipartimento di Fisica dell'Universit\`a, INFN Sezione di Pisa,
e Scuola Normale Superiore, I-56010 Pisa, Italy}
\end{center}\end{sloppypar}
\vspace{2mm}
\begin{sloppypar}
\noindent
G.A.~Blair,
G.~Cowan,
M.G.~Green,
T.~Medcalf,
J.A.~Strong,
\mbox{J.H.~von~Wimmersperg-Toeller}
\nopagebreak
\begin{center}
\parbox{15.5cm}{\sl\samepage
Department of Physics, Royal Holloway \& Bedford New College,
University of London, Surrey TW20 OEX, United Kingdom$^{10}$}
\end{center}\end{sloppypar}
\vspace{2mm}
\begin{sloppypar}
\noindent
D.R.~Botterill,
R.W.~Clifft,
T.R.~Edgecock,
P.R.~Norton,
J.C.~Thompson
\nopagebreak
\begin{center}
\parbox{15.5cm}{\sl\samepage
Particle Physics Dept., Rutherford Appleton Laboratory,
Chilton, Didcot, Oxon OX11 OQX, United Kingdom$^{10}$}
\end{center}\end{sloppypar}
\vspace{2mm}
\begin{sloppypar}
\noindent
\mbox{B.~Bloch-Devaux},
P.~Colas,
S.~Emery,
W.~Kozanecki,
E.~Lan\c{c}on,
\mbox{M.-C.~Lemaire},
E.~Locci,
P.~Perez,
J.~Rander,
\mbox{J.-F.~Renardy},
A.~Roussarie,
\mbox{J.-P.~Schuller},
J.~Schwindling,
A.~Trabelsi,$^{21}$
B.~Vallage
\nopagebreak
\begin{center}
\parbox{15.5cm}{\sl\samepage
CEA, DAPNIA/Service de Physique des Particules,
CE-Saclay, F-91191 Gif-sur-Yvette Cedex, France$^{17}$}
\end{center}\end{sloppypar}
\vspace{2mm}
\begin{sloppypar}
\noindent
S.N.~Black,
J.H.~Dann,
R.P.~Johnson,
H.Y.~Kim,
N.~Konstantinidis,
A.M.~Litke,
M.A. McNeil,
G.~Taylor
\nopagebreak
\begin{center}
\parbox{15.5cm}{\sl\samepage
Institute for Particle Physics, University of California at
Santa Cruz, Santa Cruz, CA 95064, USA$^{22}$}
\end{center}\end{sloppypar}
\vspace{2mm}
\begin{sloppypar}
\noindent
C.N.~Booth,
S.~Cartwright,
F.~Combley,
M.S.~Kelly,
M.~Lehto,
L.F.~Thompson
\nopagebreak
\begin{center}
\parbox{15.5cm}{\sl\samepage
Department of Physics, University of Sheffield, Sheffield S3 7RH,
United Kingdom$^{10}$}
\end{center}\end{sloppypar}
\vspace{2mm}
\begin{sloppypar}
\noindent
K.~Affholderbach,
A.~B\"ohrer,
S.~Brandt,
C.~Grupen,
G.~Prange
\nopagebreak
\begin{center}
\parbox{15.5cm}{\sl\samepage
Fachbereich Physik, Universit\"at Siegen, D-57068 Siegen,
 Germany$^{16}$}
\end{center}\end{sloppypar}
\vspace{2mm}
\begin{sloppypar}
\noindent
G.~Giannini,
B.~Gobbo
\nopagebreak
\begin{center}
\parbox{15.5cm}{\sl\samepage
Dipartimento di Fisica, Universit\`a di Trieste e INFN Sezione di Trieste,
I-34127 Trieste, Italy}
\end{center}\end{sloppypar}
\vspace{2mm}
\begin{sloppypar}
\noindent
J.~Rothberg,
S.~Wasserbaech
\nopagebreak
\begin{center}
\parbox{15.5cm}{\sl\samepage
Experimental Elementary Particle Physics, University of Washington, WA 98195
Seattle, U.S.A.}
\end{center}\end{sloppypar}
\vspace{2mm}
\begin{sloppypar}
\noindent
S.R.~Armstrong,
E.~Charles,
P.~Elmer,
D.P.S.~Ferguson,
Y.~Gao,
S.~Gonz\'{a}lez,
T.C.~Greening,
O.J.~Hayes,
H.~Hu,
S.~Jin,
P.A.~McNamara III,
J.M.~Nachtman,$^{2}$
J.~Nielsen,
W.~Orejudos,
Y.B.~Pan,
Y.~Saadi,
I.J.~Scott,
J.~Walsh,
Sau~Lan~Wu,
X.~Wu,
G.~Zobernig
\nopagebreak
\begin{center}
\parbox{15.5cm}{\sl\samepage
Department of Physics, University of Wisconsin, Madison, WI 53706,
USA$^{11}$}
\end{center}\end{sloppypar}
}
\footnotetext[1]{Also at CERN, 1211 Geneva 23, Switzerland.}
\footnotetext[2]{Now at University of California at Los Angeles (UCLA),
Los Angeles, CA 90024, U.S.A.}
\footnotetext[3]{Also at Dipartimento di Fisica, INFN, Sezione di Catania, 
95129 Catania, Italy.}
\footnotetext[4]{Also Istituto di Fisica Generale, Universit\`{a} di
Torino, 10125 Torino, Italy.}
\footnotetext[5]{Also Istituto di Cosmo-Geofisica del C.N.R., Torino,
Italy.}
\footnotetext[6]{Supported by the Commission of the European Communities,
contract ERBFMBICT982894.}
\footnotetext[7]{Supported by CICYT, Spain.}
\footnotetext[8]{Supported by the National Science Foundation of China.}
\footnotetext[9]{Supported by the Danish Natural Science Research Council.}
\footnotetext[10]{Supported by the UK Particle Physics and Astronomy Research
Council.}
\footnotetext[11]{Supported by the US Department of Energy, grant
DE-FG0295-ER40896.}
\footnotetext[12]{Permanent address: Kangnung National University, Kangnung, 
Korea.}
\footnotetext[13]{Supported by the US Department of Energy, contract
DE-FG05-92ER40742.}
\footnotetext[14]{Supported by the US Department of Energy, contract
DE-FC05-85ER250000.}
\footnotetext[15]{Permanent address: Universitat de Barcelona, 08208 Barcelona,
Spain.}
\footnotetext[16]{Supported by the Bundesministerium f\"ur Bildung,
Wissenschaft, Forschung und Technologie, Germany.}
\footnotetext[17]{Supported by the Direction des Sciences de la
Mati\`ere, C.E.A.}
\footnotetext[18]{Supported by Fonds zur F\"orderung der wissenschaftlichen
Forschung, Austria.}
\footnotetext[19]{Now at SAP AG, 69185 Walldorf, Germany.}
\footnotetext[20]{Now at Harvard University, Cambridge, MA 02138, U.S.A.}
\footnotetext[21]{Now at D\'epartement de Physique, Facult\'e des Sciences de Tunis, 1060 Le Belv\'ed\`ere, Tunisia.}
\footnotetext[22]{Supported by the US Department of Energy,
grant DE-FG03-92ER40689.}
\footnotetext[23]{Now at Universit\'e de Lausanne, 1015 Lausanne, Switzerland.}
\footnotetext[24]{Now at INFN Sezione de Ferrara, 44100 Ferrara, Italy.}
%
%
\setlength{\parskip}{\saveparskip}
\setlength{\textheight}{\savetextheight}
\setlength{\topmargin}{\savetopmargin}
\setlength{\textwidth}{\savetextwidth}
\setlength{\oddsidemargin}{\saveoddsidemargin}
\setlength{\topsep}{\savetopsep}
\normalsize
\newpage
\pagestyle{plain}
\setcounter{page}{1}

\normalsize

\pagenumbering{arabic}
\pagestyle{plain}

\setcounter{page}{1}

%
%
\section{Introduction}
Hadronic $\tau$ decays provide a clean environment for studying
the physics of hadrons, which are produced via $W$ exchange, 
{\it i.e.}, from the QCD vacuum. For this reason, the global properties of
hadronic systems in $\tau$ decay are described by fundamental quantities,
called spectral functions, which measure the transition probability to
create hadrons out of the vacuum as a function of the hadronic mass. Due to
their unambiguous theoretical and experimental definition, $\tau$ spectral 
functions provide information on hadron dynamics in an 
interesting mass region which is dominated by resonances and leads to the
simpler asymptotic QCD regime. The ALEPH Collaboration has already 
published analyses of the nonstrange vector~\cite{vect} and 
axial-vector~\cite{alphas} spectral functions. The nonstrange vector
spectral function can be compared to the corresponding cross
sections in $e^+e^-$ annihilation~\cite{vect} in order to test isospin
invariance of the electroweak vector current, formerly called the conserved
vector current (CVC) hypothesis. Detailed QCD studies have been 
performed using the nonstrange vector and axial-vector spectral
functions, resulting in a precise determination of the strong coupling
constant $\alpha_s(M^2_\tau)$~\cite{alphas}.
In this paper, attention is turned to the $\tau$ decays involving
kaons, which are of importance in order to address issues in both strange and
nonstrange decay dynamics~\cite{davchen,cdhtau}.       

Recently, ALEPH completed the measurement of $\tau$ decay branching ratios 
involving kaons~\cite{3prong,k0decay,1prong,keta}, as summarized in 
Table~\ref{brs1}, which allows a comprehensive 
study of the strange sector and of some aspects of 
the dynamics in the final states with a $K\overline{K}$ pair.    
These analyses involve the detection of 
charged kaons and neutral kaons ($K^0_L$, $K^0_S$).
Assuming CP invariance the $K^0_S$ and $K^0_L$ branching ratios
can be averaged. The corresponding results  
are given in the right-hand column of Table~\ref{brs1}. 
Although the strange final states are Cabibbo suppressed, 
the precision achieved in these measurements enables
many significant tests, ranging from $\mu-\tau$ lepton
universality to hadron dynamics, resonance production and QCD analyses. 
In addition, the results are checked for consistency with
isospin relations between the rates observed for
different final states.   
\begin{table}
\begin{center}
\small
\begin{tabular}{|l|c|c|c|c|}\hline\hline
Decay &$K^0$ detected &$S$ &$B~(10^{-3})$ & 
$B(K^0_L + K^0_S)~(10^{-3})$\\ \hline
$\tau^-\to K^-\nu_\tau$& $-$ &
&$ 6.96\pm0.29$ & $-$ \\
$\tau^-\to K^-\pi^0\nu_\tau$& $-$ & 
&$ 4.44\pm0.35$ & $-$\\ 
$\tau^-\to \overline{K^0}\pi^-\nu_\tau$& $K^0_L$&
&$ 9.28\pm0.56$ & \\ 
$\tau^-\to \overline{K^0}\pi^-\nu_\tau$& $K^0_S$&
&$ 8.55\pm1.34$ &\raisebox{1.2ex}[0cm][0cm]{$9.17\pm0.52$}\\ 
$\tau^-\to \overline{K^0}\pi^-\pi^0\nu_\tau$& $K^0_L$&
&$ 3.47\pm0.65$ & \\ 
$\tau^-\to \overline{K^0}\pi^-\pi^0\nu_\tau$& $K^0_S$& 
&$ 2.94\pm0.82$ &\raisebox{1.2ex}[0cm][0cm]{$3.27\pm0.51$}\\ 
$\tau^-\to K^-\pi^+\pi^-\nu_\tau$& $-$ & $-1$
&$ 2.14\pm0.47$ & $-$\\
$\tau^-\to K^-\pi^0\pi^0\nu_\tau$& $-$ &
&$ 0.56\pm0.25$ & $-$\\ 
$\tau^-\to \overline{K^0}\pi^-\pi^0\pi^0\nu_\tau$& $K^0_L$&
&$<0.66~(95\%~\mbox{C.L.})$ &\\ 
$\tau^-\to \overline{K^0}\pi^-\pi^0\pi^0\nu_\tau$& $K^0_S$&
&$ 0.58\pm0.36$ &\raisebox{1.2ex}[0cm][0cm]{$0.26\pm0.24$} \\ 
$\tau^-\to K^-\pi^0\pi^0\pi^0\nu_\tau$& $-$ &
&$ 0.37\pm0.24$ (excl. $\eta$)& $-$\\ 
$\tau^-\to K^-\pi^+\pi^-\pi^0\nu_\tau$& $-$ &
&$ 0.54\pm0.43$ (excl. $\eta$) & $-$\\ 
$\tau^-\to K^-\eta\nu_\tau$& $-$ &
&$ 0.29^{\,+0.15}_{\,-0.14}$ & $-$\\ 
$\tau^-\to K^-K^+K^-\nu_\tau$& $-$ &
&$ <0.19~(95\%~\mbox{C.L.})$ & $-$\\ \hline
$\tau^-\to K^-K^0\nu_\tau$ & $K^0_L$ &
&$ 1.62\pm0.24$ &\\ 
$\tau^-\to K^-K^0\nu_\tau$& $K^0_S$&
&$ 1.58\pm0.45$ &\raisebox{1.2ex}[0cm][0cm]{$1.61\pm0.21$}\\ 
$\tau^-\to K^-K^0\pi^0\nu_\tau$& $K^0_L$&
&$ 1.43\pm0.32$ &\\ 
$\tau^-\to K^-K^0\pi^0\nu_\tau$& $K^0_S$&
&$ 1.52\pm0.79$ &\raisebox{1.2ex}[0cm][0cm]{$1.45\pm0.30$}\\ 
$\tau^-\to K^-K^0\pi^0\pi^0\nu_\tau$& $K^0_L$&
&$ <0.18~(95\%~\mbox{C.L.})$ &\\ 
$\tau^-\to K^-K^0\pi^0\pi^0\nu_\tau$& $K^0_S$&0&
$<0.39~(95\%~\mbox{C.L.})$&\raisebox{1.2ex}[0cm][0cm]
{$<0.16~(95\%~\mbox{C.L.})$}\\ 
$\tau^-\to K^0_SK^0_L\pi^-\nu_\tau$& $-$ &
&$ 1.01\pm0.26$ &\\ 
$\tau^-\to K^0_SK^0_S\pi^-\nu_\tau$& $-$ &
&$ 0.26\pm0.12$ &\raisebox{1.2ex}[0cm][0cm]{$1.53\pm0.35$} \\
$\tau^-\to K^0_SK^0_L\pi^-\pi^0\nu_\tau$& $-$ &
&$ 0.31\pm0.12$ &\\ 
$\tau^-\to K^0_SK^0_S\pi^-\pi^0\nu_\tau$& $-$ &
&$ <0.20$ (95$\%$ C.L.)&\raisebox{1.2ex}[0cm][0cm]{$0.31\pm0.23$}\\ 
$\tau^-\to K^-K^+\pi^-\nu_\tau$& $-$ &
&$ 1.63\pm0.27$ & $-$\\ 
$\tau^-\to K^-K^+\pi^-\pi^0\nu_\tau$& $-$ &
&$ 0.75\pm0.33$ & $-$\\ \hline
$\tau^-\to K^0h^-h^+h^-\nu_\tau$& $K^0_S$& mixed
& $0.23\pm0.20$ & $0.23\pm0.20$\\ \hline\hline
\end{tabular}
\caption{\it Summary of branching ratios for 
$\tau$ decays involving kaons from ALEPH data.
Channels with neutral kaons are measured separately using their 
$K^0_L$ and $K^0_S$ components.
Modes are split into the strange $(S=-1)$ and nonstrange $(S=0)$
sectors of $\tau$ decays. Due to the limitations of statistics and 
particle identification, the net strangeness of the $K^0h^-h^+h^-$ mode 
has not been determined (h stands for $\pi$ or $K$).
The last column presents the average of 
$K^0_S$ and $K^0_L$ results for final states with $\overline{K^0}$ and
$K^-K^0$, and the sum of all components for final states with 
$K^0\overline{K^0}$.}
\label{brs1}
\end{center}
\end{table}

One of the outstanding issues in the nonstrange sector is the vector
and axial-vector composition of the $K\overline{K}\pi$ final state. This was
a limitation in the QCD analysis of Ref.~\cite{alphas}.
New information on this point is obtained from the ALEPH measurement
of all relevant decay modes\footnote{Throughout this paper, charge conjugate 
states are implied.} $K^0_SK^0_L\pi^-$, $K^0_SK^0_S\pi^-$, $K^+K^-\pi^-$ and 
$K^0K^-\pi^0$.
In addition, the availability of the corresponding isovector cross 
section measured in $e^+e^-$ annihilation and assuming CVC allows an 
independent determination of the vector fraction in the 
$(K\overline{K}\pi)^-\nu_\tau$ channel, with a significant
improvement in accuracy.

Because G-parity cannot be defined for the strange hadronic states, 
it is difficult to separate vector and axial-vector contributions. 
Furthermore, the relatively low statistics do not permit different 
spin-parity states to be unfolded from the overall decay distributions.
Experimentally, the $\overline{K}\pi\pi$ system 
is the most complex one to deal with, since it contains contributions from
two axial-vector mesons, $K_1(1270)$ and $K_1(1400)$, and 
one vector meson, $K^{*}(1410)$. However,
the $K^{*}(1410)$ state can be measured in the 
$(\overline{K}\pi)^-\nu_\tau$ channel by its interference with 
the dominant $K^*(892)$ production, analogously to the $\rho(770)-\rho(1450)$ 
situation in the $\pi\pi^0$ final states~\cite{vect}. 
Unlike the $K_1(1400)$ and $K^{*}(1410)$, 
the axial-vector $K_1(1270)$ decays significantly into $K\rho$.
And so fits to the invariant $\pi\pi$ mass 
distributions~\cite{3prong,k0decay,1prong}, which measure 
the $\overline{K}\rho$ fraction, determine 
the total $K_1(1270)$ contribution.
A fit to the invariant $\overline{K}\pi\pi$ mass spectrum 
provides additional information on the resonance content. 
It is important to separate the vector and axial-vector 
contributions in the strange sector of $\tau$ decays to obtain 
information on the size of the nonperturbative QCD part in the $\tau$ 
hadronic width. This has been observed to be very small in
the nonstrange case~\cite{alphas}.

Independently of the resonance structure, the total strange spectral function
is determined, without separating vector and axial-vector contributions.
Similar to the nonstrange sector, the $\tau$ strange
spectral functions are key ingredients for QCD studies and provide the 
possibility to follow the results as a function of 
a variable ``$\tau$ mass''~\cite{alphas}.

One of the free parameters of the Standard Model, 
the strange quark mass $m_s$, appears in
many phenomenological calculations, such as in the prediction of 
the CP-violating kaon parameters $\epsilon '/\epsilon$~\cite{buras}.
The strange $\tau$ decay rate is sensitive to $m_s$~\cite{bnp}
which can be experimentally determined.
A breaking of chiral symmetry, induced by the relatively large 
strange quark mass, introduces a mass dependence into 
the perturbative QCD prediction
of the total strange hadronic width of the $\tau$. 
Using the total rates and moments of the respective spectral 
functions, a combination of strange and nonstrange modes can be found
which cancels the dominant massless perturbative contribution, 
enabling the strange quark mass to be derived by means of a combined fit 
with nonperturbative contributions. An error in the theoretical 
calculation~\cite{chetkwiat} of the second-order perturbative term
was recently pointed out~\cite{maltman}. After correction, a problematic
convergence behaviour of the perturbative series is observed   
~\cite{maltman,pichprad,kuhnchet}, leading to a larger theoretical 
uncertainty on $m_s$ than had previously been assumed and to a bias 
in the central value obtained~\cite{davchen}. In this analysis, particular
attention is devoted to this question and a new method is 
introduced~\cite{cdh} to avoid the convergence problem.

\section{$\mu-\tau$ lepton universality}

Lepton universality requires that the $W^-$ boson coupling to any
lepton pair $l\bar{\nu}_l$ be of the same strength $g_l$; this hypothesis is in
agreement with studies in the leptonic and hadronic sectors~\cite{rolandi}. 
Lepton universality can be tested in the strange sector 
by comparing the decay widths of 
$\tau^-\to K^-\nu_{\tau}$ and $K^-\to \mu^-\bar{\nu}_{\mu}$. The following 
ratio, in which the common CKM matrix element and the decay constant $f_K$
both cancel, is precisely predicted by theory: 
\begin{eqnarray}
R_{\tau/K}&\equiv&\frac{\Gamma(\tau^-\to K^-\nu_\tau)}
            {\Gamma(K^-\to\mu^-\bar{\nu}_\mu)} \nonumber \\
          &\equiv&\frac{B(\tau^-\to K^-\nu_{\tau})\tau_K}
            {B(K^-\to\mu^-\bar{\nu}_\mu)\tau_\tau} \nonumber \\
          &=&\frac{1}{2}\frac{g_\tau^2}{g_\mu^2}\frac{M_\tau^3}{M_K M_\mu^2}
            \frac{(1-M_K^2/M_\tau^2)^2}{(1-M_K^2/M_\tau^2)^2}
            (1+\delta R_{\tau/K})~,
\label{rtk}
\end{eqnarray}
with the radiative correction 
$\delta R_{\tau/K}=(0.90\pm0.22)\%$~\cite{decker}, 
$B(K^-\to\mu^-\nu_{\tau})=B_K=(63.51\pm0.18)\%$, 
$\tau_K=(1.2386\pm0.0024)\times10^{-8}$ s,
and $\tau_{\tau}=(290\pm1.2)\times10^{-15}$ s~\cite{pdg98}. Assuming 
$\mu-\tau$ universality ($g_\tau=g_\mu$), Eq.~(\ref{rtk}) predicts
\begin{equation}
B(\tau^-\to K^-\nu_{\tau})
=(7.14\pm0.02_{th}\pm0.02_{B_K}\pm0.03_{\tau_\tau})\times10^{-3}~,
\label{brkaonthe}
\end{equation}
where only sizeable contributions to the errors are given
(the theoretical uncertainty is from radiative corrections).

The comparison of (\ref{brkaonthe}) with the ALEPH measurement 
\begin{equation}
B(\tau^-\to K^-\nu_{\tau})=(6.96\pm0.25\pm0.14)\times10^{-3}
\label{brkaonexp}
\end{equation}
yields
\begin{equation}
g_{\tau}/g_{\mu}=0.987\pm0.021,
\end{equation} 
which agrees with the universality assumption within one standard deviation.

The measurement~(\ref{brkaonexp}) can also be exploited to determine
$f_K$ if lepton universality is assumed to hold. 
One obtains\footnote{The decay
constant $f_P$ should be scaled by a factor
of $\sqrt{2}$ to compare with the value given in Ref.~\cite{pdg98}.}
\begin{equation}
f_K=\frac{1}{G_F|V_{us}|}\left(1-\frac{M_K^2}{M_\tau^2}\right)^{-1}
M_\tau^{-\frac{3}{2}}\sqrt{\frac{B(\tau^-\to K^-\nu_\tau)}{\tau_\tau}}~,
\end{equation}
where $G_F=(1.16639\pm0.00002)\times 10^{-5}$ GeV$^{-2}$~\cite{pdg98} is 
the Fermi coupling constant,
and radiative corrections are ignored. This results in
\begin{equation} 
f_K=(111.5\pm2.3\pm0.9)~\mbox{MeV}~,
\label{fkdecont}
\end{equation}
where the first error is from the uncertainty on
$B(\tau^-\to K^-\nu_\tau)$ and the second from $V_{us}$.

\section{Study of resonance structure and dynamics}

The decay dynamics of $\tau$ decays involving kaons are 
investigated via the invariant mass spectra. Due to the small statistics,
decay modes other than $(\overline{K}\pi)^-$,
$(\overline{K}\pi\pi)^-$ and $(K\overline{K}\pi)^-$ are not discussed here.
Even for these, only a qualitative approach based on simple
descriptions can be considered.

\subsection{Mass spectrum for $\tau^-\to(\overline{K}\pi)^-\nu_\tau$}

Three measured modes, $K^-\pi^0$, $K^0_S\pi^-$ and $K^0_L\pi^-$,
are used to study the invariant mass of the $\overline{K}\pi$ system. 
The $K^*(892)^-$ dominance is well established in 
all the cases~\cite{k0decay,1prong}. 
For the first two modes, where the mass resolution is good enough, 
some excess is observed at
large mass over and above the expected $K^*(892)^-$ tail.
This could arise from a possible $K^*(1410)$ vector state contribution, 
expected to be dominated by $K^*(892)-K^*(1410)$
interference as already observed for $\rho(770)-\rho(1450)$ in the 
$\pi^-\pi^0$ system~\cite{vect}. Theoretically, the decay rate for 
$\tau^-\to(\overline{K}\pi)^-\nu_\tau$ relative to the electronic width
is expressed as~\cite{thacker,fink}
\begin{eqnarray}
\Gamma_{K\pi}/\Gamma_e
&\sim &\int\frac{ds}{M^2_\tau}
      \left(1-\frac{s}{M^2_\tau}\right)^2
      \left(1+2\frac{s}{M^2_\tau}\right)\nonumber \\
&   & \times\left(s-M_+^2\right)^{3/2}
\left(s-M_-^2\right)^{3/2}s^{-3}|f(s)|^2~,
\end{eqnarray}
where $M_{\pm}=M_K\pm M_\pi$ and the form factor includes both 
$K^*(892)$ and $K^*(1410)$ contributions, namely
\begin{equation}
f(s) = \frac{1}{1+\beta}[BW_{K^*(892)}(s) + \beta BW_{K^*(1410)}(s)]~,
\end{equation}
where $\beta$ is assumed to be real~\cite{fink}.
The two Breit-Wigner propagators have the form 
\begin{equation}
BW(s)=\frac{M^2_0}{M^2_0-s-i\sqrt{s}\Gamma(s)}~,     
\end{equation}
with an energy dependent width $\Gamma(s)$
\begin{equation}
\Gamma(s)=\Gamma_0\frac{M_0^2}{s}\left(\frac{p(s)}{p(M^2_0)}\right)^{2l+1}~,
\end{equation}
\begin{equation}
p(s)=\frac{1}{2\sqrt{s}}\sqrt{(s-M_+^2)(s-M_-^2)}~,
\end{equation}
where $l$ is the $K-\pi$ angular momentum, $i.e.$, $l=1$
for p wave. 

\begin{figure}
\begin{center}
\vspace{-2.0cm}
\epsfxsize 10cm
\epsffile{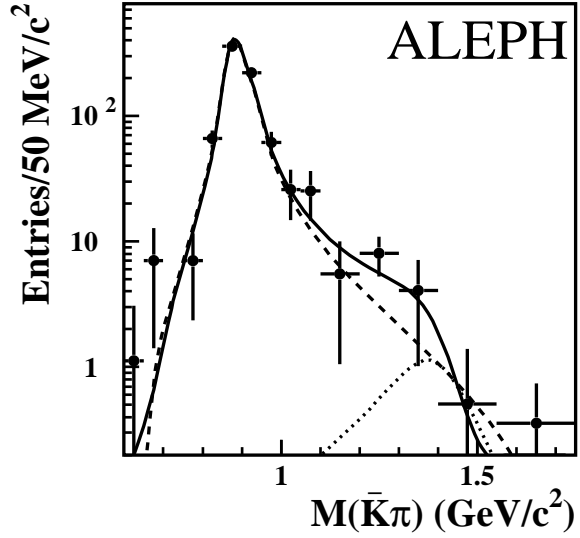}
\end{center}
\vspace{-0.5cm}
\caption{\it The invariant mass spectrum for 
$\tau^-\to K^-\pi^0\nu_\tau$ after background subtraction,
bin migration and acceptance corrections.
The fit is described by the solid curve, taking into account both
$K^*(892)$ and $K^*(1410)$ contributions with interference.
The dashed and dotted curves show the $K^*(892)$ and $K^*(1410)$
parts.}
\label{mkstar}
\end{figure}

The $K^*(1410)$ search is performed using only the $K^-\pi^0$
channel as this has the lowest background in the interference
region and the best mass resolution. Figure~\ref{mkstar} shows 
the mass spectrum after background subtraction and bin by bin
correction for both resolution and acceptance.
The fits using the above parametrization are listed in Table~\ref{kstarfit}.
If the mass and width of the $K^*(892)$ are left free (Fit 1), the
fitted values are found to be consistent with the world
average~\cite{pdg98}. With the fixed $K^*$ parameters, the final value
is obtained (Fit 2) using the $K^-\pi^0$ data, where the 
systematic uncertainty from the non-$K\pi$ events is negligible.  
A fit to the combined $K^-\pi^0$ and $K^0_S\pi^-$
invariant mass distributions is also given (Fit 3).

The measurement of $\beta=-0.11\pm0.05$ is, as expected from 
$SU(3)$ flavour symmetry, in agreement with 
the $\rho(1450)$ contribution 
($\beta=-0.094\pm0.007$ from the ALEPH measurement
of the decay $\tau^-\to \pi^-\pi^0\nu_\tau$~\cite{vect}).
Low statistics do not allow much more information to be derived
from the fit; however, leaving the phase of $\beta$ as a free
parameter yields a value $(180 \pm 55)^\circ$,
consistent with the assumption made.

The relative $K^*(892)$ and $K^*(1410)$ ``diagonal'' contributions to the 
$\tau^-\to(\overline{K}\pi)^-\nu_\tau$ branching ratio 
determined from the value of $\beta$ are found to be 
$(97.5^{+1.0}_{-1.4})\%$ and $(0.8^{+0.7}_{-0.5})\%$, respectively,
the interference contributing $(1.7^{+0.7}_{-0.5})\%$
for the remainder. According to Ref.~\cite{pdg98} the 
decay fraction of $K^*(1410)$ into $\overline{K}\pi$ is 
$(6.6\pm1.3)\%$, and for this analysis the rest is assumed to decay into 
$\overline{K}\pi\pi$ final states. After correction for the $\tau$ 
kinematic factor (see Eq.~(\ref{spect})), this fraction 
becomes $(7.2\pm1.4)\%$ for $K^*(1410)$ produced in $\tau$ decays
and leads to 
\begin{equation}
B(\tau^-\to K^*(1410)^-\nu_\tau)
=(1.5^{\,+1.4}_{\,-1.0})\times10^{-3}~,
\end{equation} 
using the $\overline{K}\pi$ branching fraction from Table~\ref{brs1}.

In this analysis, a contribution from the scalar $K^*_0(1430)$ 
has been ignored.
No evidence for this state is seen in the $\overline{K}\pi$ mass plot.
Nevertheless, a fit with this contribution was 
performed using the resonance parameters from Ref.~\cite{pdg98}. 
A $95\%$ confidence level upper limit of $0.5\times 10^{-3}$ 
is derived for the corresponding branching ratio. 
\begin{table}
\begin{center}
\begin{tabular}{|l|c|c|c|}\hline\hline
Parameter             &  Fit 1 $(K^-\pi^0)$ & Fit 2 $(K^-\pi^0)$       
& Fit 3 $(K^-\pi^0+K^0_S\pi^-)$         \\ \hline 
$M_{K^*(892)}$ (GeV/$c^2$)             & $0.895\pm0.002$ & 0.892  & 0.892 \\ 
$\Gamma_{K^*(892)}$ (GeV/$c^2$)        & $0.055\pm0.008$ & 0.050  & 0.050 \\ 
$\beta$               & $-0.06\pm0.07 $ & $-0.11\pm0.05$ & $-0.12\pm0.04$\\ 
$M_{K^*(1410)}$ (GeV/$c^2$)          & 1.412   & 1.412  & 1.412 \\ 
$\Gamma_{K^*(1410)}$ (GeV/$c^2$)     & 0.227   & 0.227  & 0.227 \\ \hline
$\chi^2$/ndf          & 13.0/19         & 16.0/21        & 19.8/21\\ 
\hline\hline
\end{tabular}
\caption{\it Fit results for $\tau^-\to K^-\pi^0\nu_\tau$ and 
$\tau^-\to K^0_S\pi^-\nu_\tau$ decays. The errors given are statistical only.
Because the background of $\tau^-\to K^-K^0\nu_\tau$ dominates 
over the higher $K^0_S\pi^-$ invariant mass tail,
the background subtraction relies on the model prediction~\cite{fink} and
introduces additional systematic uncertainties in the $K^0_S\pi^-$ system.
The fixed values for the masses and widths of the $K^*$ resonances are 
taken from Ref.~\cite{pdg98}.}
\label{kstarfit}
\end{center}
\end{table}

\subsection{Mass spectra for $\tau^-\to(\overline{K}\pi\pi)^-\nu_\tau$}

The $(\overline{K}\pi\pi)^-$ decay occurs in the
four final states $K^-\pi^+\pi^-$, $K^0_S\pi^-\pi^0$, $K^0_L\pi^-\pi^0$ and
$K^-\pi^0\pi^0$. Its invariant mass spectrum as well as the spectra for the
$\overline{K}\pi$ and $\pi\pi$ subsystems reveal details of the 
underlying decay dynamics.

\subsubsection{Contribution from $K^*(1410)$}

The fitted contribution of $K^*(1410)$ found in the previous section
has consequences for the $(\overline{K}\pi\pi)^-$ final state.
The fraction $(92.8\pm1.4)\%$ of $K^*(1410)$ decaying into 
$\overline{K}\pi\pi$ is used to estimate the vector $\overline{K}\pi\pi$
component, while the remainder is attributed to
the axial-vector contribution dominated by $K_1(1270)$ and $K_1(1400)$.
For the vector current part, one obtains
\begin{equation}
B(\tau^-\to K^*(1410)^-\nu_\tau\to (\overline{K}\pi\pi)^-\nu_\tau)=
(1.4^{\,+1.3}_{\,-0.9}~^{\,+0.0}_{\,-0.4})\times10^{-3}~,
\label{vk1}
\end{equation}
where the first uncertainty comes from the fit to
the $\overline{K}\pi$ invariant 
mass, while the second uncertainty arises from the possibility for 
the $K^*(1410)$ to decay into $K\eta$. After subtracting this
vector component, the axial-vector contribution
to the $\overline{K}\pi\pi$ decay modes is
\begin{equation}
B_A(\tau^-\to (\overline{K}\pi\pi)^-\nu_\tau)
=(4.6^{\,+1.2}_{\,-1.5})\times 10^{-3}~.
\label{bkppa}
\end{equation}

\subsubsection{$\overline{K}\rho$ and $\overline{K}^*\pi$ fractions}

The $\overline{K}\pi$ and $\pi\pi$ mass spectra are 
used to search for the intermediate
states $\overline{K}^*\pi$ and $\overline{K}\rho$. 
The $\overline{K}\rho$ fraction can reveal the presence of 
the axial vector $K_1(1270)$ resonance which decays
about $50\%$ of the time to $K\rho$~\cite{pdg98}, 
unlike the $K_1(1400)$ which decays
almost purely to $\overline{K}^*\pi$. For practical reasons related to the
shape of the different backgrounds, the most efficient way 
for separating the $\overline{K}^*\pi$ and 
$\overline{K}\rho$ intermediate states 
is to fit the $\pi\pi$ invariant mass spectra where
background and non-$\overline{K}\rho$ contributions 
are essentially located at lower masses. 

\begin{figure}
\begin{center}
\vspace{-2.0cm}
\epsfxsize 10cm
\epsffile{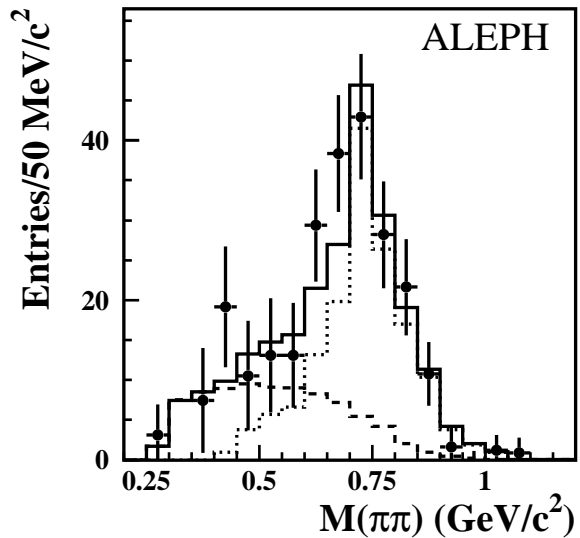}
\end{center}
\vspace{-0.5cm}
\caption{\it The $\pi\pi$ invariant mass spectrum for 
$\tau^-\to (\overline{K}\pi\pi)^-\nu_\tau$ after background subtraction.
The overall fit, the expected $\rho$ signal and $\overline{K}^*\pi$ 
reflection are shown in solid, dotted and dashed histograms.} 
\label{mpipi}
\end{figure}

No excess in the $\rho$ mass region is found in the $K^-\pi^0\pi^0$
mode, as expected, while a $\rho$ signal is 
observed in the $K^-\pi^+\pi^-$ mode~\cite{3prong}
and also in both the $K^0_S\pi^-\pi^0$~\cite{k0decay} and 
$K^0_L\pi^-\pi^0$ modes~\cite{1prong}. With the assumption that
the $\overline{K}\pi\pi$ final states proceed only via 
an incoherent superposition of
the intermediate states 
$\overline{K}\rho$ and $\overline{K}^*\pi$, a sum of
a $\rho$ Breit-Wigner signal with the $\tau$ kinematic factor 
and the shape of the $\overline{K}^*\pi$ reflections 
obtained from simulation is used to fit the $\pi\pi$ mass spectrum after 
subtracting the non-$\overline{K}\pi\pi$ background. It yields the fractions  
$f_{K^-\rho^0}=(39\pm14)\%$ for the $K^-\pi^+\pi^-$ 
mode\footnote{This value differs slightly
from the published result ($35 \pm 11$) in Ref.~\cite{3prong}
which was obtained with a $\rho$ lineshape inadvertently
not corrected for the $\tau$ kinematic factor.} and 
$f_{\overline{K^0}\rho^-}=(74\pm13)\%$ for the 
$\overline{K^0}\pi^-\pi^0$ mode.
The sum of the fits to the $\pi\pi$ invariant mass spectra is shown in
Fig.~\ref{mpipi}.
These results indicate that the $(\overline{K}\rho)^-$ intermediate state 
plays an important role in the $\overline{K}\pi\pi$ mode. The smaller value in 
the $K^-\pi^+\pi^-$ mode is understood from isospin considerations 
as discussed in Section 5. The sum of decays involved in the
$\overline{K}\rho$ intermediate state gives 
$B(\tau^-\to(\overline{K}\rho)^-\nu_\tau)=(3.25\pm0.67)\times10^{-3}$.

The branching fraction of the $K_1(1270)$ into $K\rho$
is taken from Ref.~\cite{pdg98} and becomes 
$(52\pm7)\%$, when taking into account the $\tau$
kinematic factor. A first estimate of the branching ratio 
to the $K_1(1270)$ is therefore derived to be
\begin{equation}
B(\tau^-\to K^-_1(1270)\nu_\tau)=(6.3\pm1.5)\times 10^{-3}
~~~(\overline{K}\rho~\mbox{fractions})~,
\label{b1k1}
\end{equation}
where all the $\overline{K}\rho$ states are assumed to 
originate from $K_1(1270)$ decay only.

\subsubsection{Fit to $\overline{K}\pi\pi$ resonances}

\begin{figure}
\begin{center}
\vspace{-2.0cm}
\epsfxsize 10cm
\epsffile{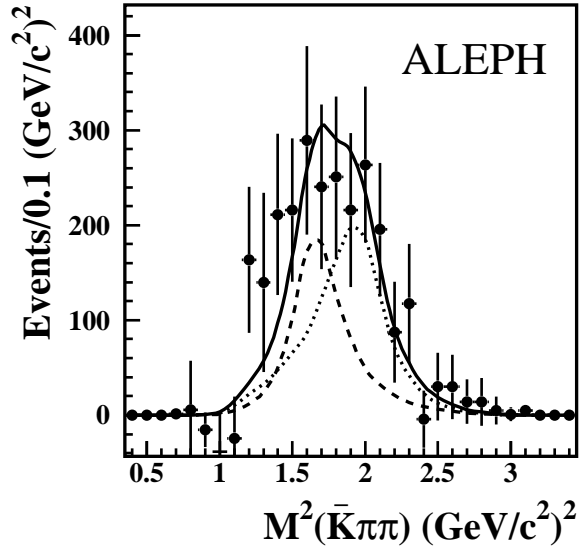}
\end{center}
\vspace{-0.5cm}
\caption{\it The invariant mass-squared distribution for 
$\tau^-\to (\overline{K}\pi\pi)^-\nu_\tau$ after efficiency corrections 
for the different final states and background subtraction.
Data are shown by the dots with error bars. 
The fit is described by the solid curve, while the expected  
$K_1(1270)$ and $K_1(1400)$ contributions are separately depicted by
the dashed and the dotted curves.}
\label{mkpp}
\end{figure}

The description of the $\overline{K}\pi\pi$ mass spectrum in terms of the 
resonant states $K_1(1270)$, $K_1(1400)$ and $K^*(1410)$ is
rather complicated because several intermediate states are involved,
as discussed in Ref.~\cite{slac,accmor}. Therefore,
the shape of the respective mass distributions 
is obtained from the Monte Carlo simulation, 
taking into account all the known intermediate states.
The resonance parameters for the $K_1(1270)$, $K_1(1400)$ and $K^*(1410)$ are
fixed at 1.273 GeV/$c^2$, 1.402 GeV/$c^2$ and 1.412 GeV/$c^2$ for the
masses and at 0.15 GeV/$c^2$, 0.174 GeV$c^2$ and 0.227 GeV/$c^2$ for
the widths. The chosen width for the $K_1(1270)$ is different 
from the value in Ref.~\cite{pdg98},
because of the parametrization adopted: it should be interpreted as an 
effective width, resulting from the opening of the different decay 
channels. It has been checked that the resulting line shape agrees 
well with the existing data~\cite{slac,accmor}.
However as the value for the width can have a strong effect 
on the relative fractions in the 
$K_1(1270)$ decay, an uncertainty of $\pm0.05$ GeV/$c^2$ is
assigned to it.
The mass resolution and the statistics are not sufficient to separate
the $K_1(1400)$ and $K^*(1410)$ states, and so an effective resonance 
is used instead, averaging the parameters of the two states.
 To obtain the shapes including all the intermediate states,
the $K_1(1270)$ decays into $K\rho$, $K^*(892)\pi$,
$K^*(1430)\pi$ and $Kf_0(1370)$ are generated 
separately and then combined
according to the relative fractions given in Ref.~\cite{pdg98}. 
Since the $\tau$ kinematic factor modifies 
the branching ratios, the fractions for the different final states 
are recomputed for $K_1(1270)$ production 
in $\tau$ decays, yielding 
$(52\pm7)\%$, $(25\pm8)\%$, $(8\pm1)\%$, $(14\pm3)\%$ and $(1\pm1)\%$ for
the $K\rho$, $K^*(892)\pi$, $K^*(1430)\pi$ $K\omega$ 
and $Kf_0$ final states, respectively. The fit is  
shown in Fig.~\ref{mkpp}, giving $\chi^2/ndf=14.1/22$ and a fraction 
\begin{equation}
f_{K_1(1270)}=(41\pm19\pm15)\% ~,~
\end{equation}
where the first error is of statistical origin and the second comes
from the $K_1(1270)$ width. Because the
total branching ratio for the $\overline{K}\pi\pi$ modes is 
$B(\tau^-\to (\overline{K}\pi\pi)^-\nu_\tau)=(5.97\pm0.72)\times 10^{-3}$,
the expected branching ratio for $K_1(1270)$ is 
\begin{equation}
B(\tau^-\to K^-_1(1270)\nu_\tau)=(2.9\pm1.7)\times 10^{-3}
~~~(\overline{K}\pi\pi~\mbox{fit})~,~
\label{b2k1}
\end{equation}
taking into account the fact that 
$(86\pm3)\%$ of the $K_1(1270)$'s decay into $K\pi\pi$ final
states. 

The values (\ref{b1k1}) and~(\ref{b2k1}) obtained from two 
independent estimates are in fair agreement and their combination yields
\begin{equation}
B(\tau^-\to K^-_1(1270)\nu_\tau)=(4.8\pm1.1)\times 10^{-3}
\label{k1270}
~~~(\mbox{combined})~.
\end{equation}
From this value, the $\tau$ decay into $K^-\omega$ through $K^-_1(1270)$ is  
estimated (using the branching ratio in Ref~\cite{pdg98}) 
to be $(0.67\pm0.21)\times 10^{-3}$, indicating an axial-vector contribution
of $B_A(\tau^-\to K^-\pi^+\pi^-\pi^0\nu_\tau)=(0.59\pm0.19)\times 10^{-3}$ in 
the measured $\overline{K}\pi\pi\pi$ decay modes. 
Comparing the results in Eq.~(\ref{bkppa}) and Eq.~(\ref{k1270}),
one obtains a branching ratio
$B(\tau^-\to K^-_1(1400)\nu_\tau)=(0.5\pm1.7)\times 10^{-3}$, which
is somewhat reduced in comparison with that of $K_1(1270)$.

\subsection{Mass spectra for $\tau^-\to(K\overline{K}\pi)^-\nu_\tau$}

Almost all possible final states for 
$(K\overline{K}\pi)^-$ are studied by ALEPH (Table~\ref{brs1}):
$K^+K^-\pi^-$, $K^0_SK^0_S\pi^-$, 
$K^0_SK^0_L\pi^-$, $K^-K^0_S\pi^0$ and $K^-K^0_L\pi^0$.
The $K^+\pi^-$ mass plot in the $K^+K^-\pi^-$ mode shows a clear 
$K^*(892)^0$ signal, and a fit yields a fraction of $(87\pm13)\%$
for the $K^*K^-$ component~\cite{3prong}.
A strong $K^*(892)^-$ signal is also found in the $K^-K^0_L\pi^0$ mode with
an estimated lower limit of $86\%$ at $90\%$ confidence level for the 
$\overline{K}^*(K^*)$ component~\cite{1prong}, 
confirming the conclusion of $K^*$ dominance in 
the $K\overline{K}\pi$ mode. 

\begin{figure}
\begin{center}
\vspace{-2.0cm}
\epsfxsize 10cm
\epsffile{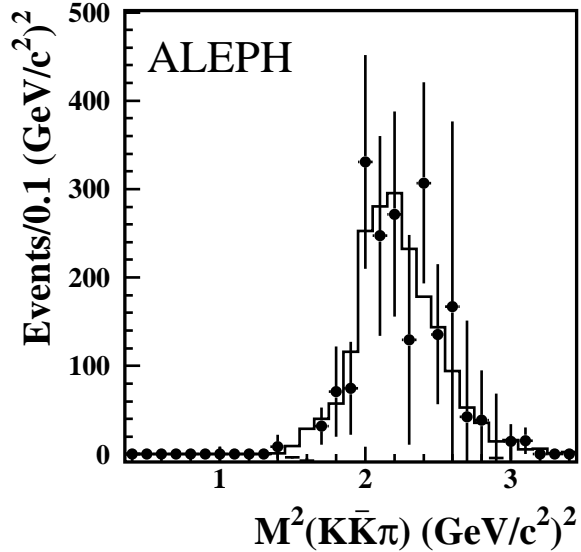}
\end{center}
\vspace{-0.5cm}
\caption{\it The invariant mass-squared distribution for 
$\tau^-\to (K\overline{K}\pi)^-\nu_\tau$ after efficiency corrections for the 
different final states and background subtraction.
Data are shown by the dots with error bars. The histogram represents the
model expectation~\cite{dms}.}
\label{mkkp}
\end{figure}

Both the $K^+K^-\pi^-$ and $K^-K^0_S\pi^0$ modes are used to measure 
the total $K\overline{K}\pi$ invariant mass distribution, scaling
their rates by a factor of two. Figure~\ref{mkkp} shows the 
combined $K\overline{K}\pi$ invariant
mass-squared distribution after correcting for resolution and efficiency.
The Monte Carlo expectation~\cite{dms}, based on $a_1$ dominance,
is observed to be consistent with the data.

\section{Tests of isospin invariance}

Isospin relations relevant to $\tau$ decays into kaons have been 
established in Ref.~\cite{gilman,rouge,rouge1}. 
The set of results obtained by ALEPH allows for
a complete test of these relations. Apart from testing isospin invariance
in the hadronic states accessible in $\tau$ decays, this procedure can be 
helpful as a consistency check among the different final states, 
regardless of the decay dynamics.  

\subsection{The decay $\tau^-\to(\overline{K}\pi)^-\nu_\tau$}

\begin{figure}
\begin{center}
\vspace{-2.0cm}
\epsfxsize 10cm
\epsffile{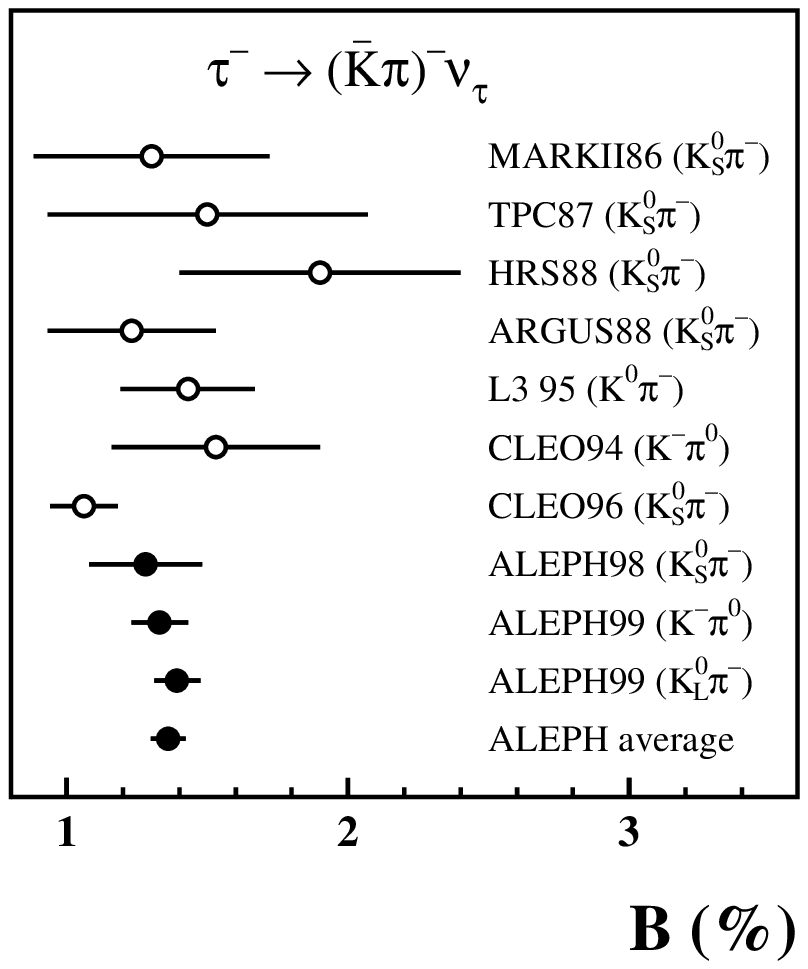}
\end{center}
\vspace{-0.5cm}
\caption{\it The published branching ratios for 
$\tau^-\to (\overline{K}\pi)^-\nu_\tau$~\cite{k0decay,1prong,kpibrs}, 
using all three measured modes and assuming $I=1/2$.}
\label{brkstar}
\end{figure}

The branching ratios of the three final states of the decay 
$\tau^-\to(\overline{K}\pi)^-\nu_\tau$, respectively,
$K^-\pi^0$, $K^0_S\pi^-$ and $K^0_L\pi^-$, are presented 
in Table~\ref{brs1}.
A study of the corresponding invariant mass spectra shows that the decay is
dominated by $K^*(892)$, with a small component of $K^*(1410)$
visible through its interference with $K^*(892)$ as presented in Section~3.1.
Isospin symmetry for an $I=1/2$ state, together with
the assumption of equal $K^0_S$ and $K^0_L$ contributions, constraints the 
rates for the three final states
to be equal up to small mass corrections. Taking this into account,
the above three measurements are 
well consistent with isospin symmetry, giving a $\chi^2/ndf=0.9/2$. 
Using isospin symmetry as a constraint yields the total branching ratio 
\begin{equation}
B(\tau^-\to (\overline{K}\pi)^-\nu_\tau)=(13.60\pm0.62)\times 10^{-3}~.
\end{equation} 
A comparison between this analysis and the other measurements is shown
in Fig.~\ref{brkstar}.

From the fit performed in Section~3.1, the branching ratio for
$K^*(892)^-$ production is found to be
\begin{equation}
B(\tau^-\to K^*(892)^-\nu_\tau)=(13.26\pm0.63)\times 10^{-3}~.
\end{equation} 

\subsection{The decay $\tau^-\to(\overline{K}\pi\pi)^-\nu_\tau$}

\begin{figure}
\begin{center}
\vspace{-2.0cm}
\epsfxsize 10cm
\epsffile{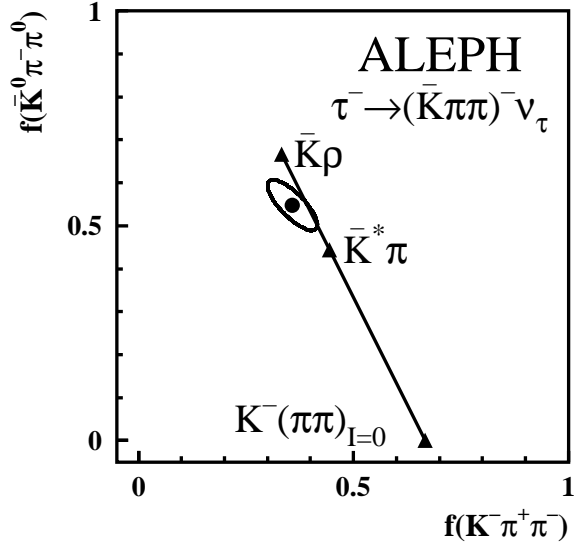}
\end{center}
\vspace{-0.5cm}
\caption{\it The decay fractions for $K^-\pi^+\pi^-\nu_\tau$ and
$\overline{K^0}\pi^0\pi^-\nu_\tau$ given by the 
standard error ellipse (39$\%$ probability).
The isospin constraint is shown by the straight line. The triangles
indicate possible intermediate states dominated by $\overline{K}^*\pi$,
$\overline{K}\rho$ or $K^-(\pi\pi)_{I=0}$.}
\label{fkpipi}
\end{figure}

For a total isospin $I=1/2$ and the two possible isospin values 
$I_{\pi\pi}=0,1$ for the $\pi\pi$ system, a relation can be
established among the branching ratios of the $\overline{K^0}\pi^-\pi^0$,
$K^-\pi^+\pi^-$ and $K^-\pi^0\pi^0$ modes~\cite{rouge}:
\begin{equation}
B(\tau^-\to K^-\pi^+\pi^-\nu_\tau)
=\frac{1}{2}B(\tau^-\to \overline{K^0}\pi^-\pi^0\nu_\tau)
+2B(\tau^-\to K^-\pi^0\pi^0\nu_\tau)~.
\label{kpipi}
\end{equation}
In order to examine to what extent 
the above isospin constraint is satisfied 
by experiment, it is convenient to use the 
relative fractions of the total $\overline{K}\pi\pi$ 
branching ratio for each individual mode. This gives
$f(\overline{K^0}\pi^-\pi^0)=0.55\pm0.06$,
$f(K^-\pi^+\pi^-)=0.36\pm0.06$, and $f(K^-\pi^0\pi^0)=0.09\pm0.04$ 
from the ALEPH measurements with
a correlation coefficient of $-0.8$ between 
$f(\overline{K^0}\pi^-\pi^0)$ and $f(K^-\pi^+\pi^-)$. The corresponding 
error ellipse is given in Fig.~\ref{fkpipi}, showing that
the isospin relation~(\ref{kpipi}) is satisfied within one standard deviation.

The isospin relations can be tested in more detail if 
the different intermediate states are separated.
Three possible intermediate states
occurring in the $(\overline{K}\pi\pi)^-$ final state are considered: 
$\overline{K}\rho$ and $K^-(\pi\pi)_{I=0}$ 
corresponding to $I_{\pi\pi}=1$ and 0, 
respectively, and $\overline{K}^*\pi$, with the
expected relative rates in the different final states as given
in Table~\ref{isokpp}.  
It turns out that a pure $K^-(\pi\pi)_{I=0}$ is excluded by
the measurement, while a combination of $\overline{K}\rho$ and  
$\overline{K}^*\pi$ can fit the data well. 
This is in agreement with the 
findings of the resonance structure analysis performed in Section~3.2.2. 
A further check of the isospin symmetry is achieved taking advantage of  
the $\overline{K}\rho$ and $\overline{K}^*\pi$ separation provided by
the $\rho$ resonance fits performed in the two relevant channels. 
The results given in Table~\ref{krho} are in agreement with the expectations. 
\begin{table}
\begin{center}
\begin{tabular}{|c|c|c|c|c|}\hline\hline
Mode &$K^-\pi^+\pi^-$ & $\overline{K^0}\pi^-\pi^0$ 
& $K^-\pi^0\pi^0$ & $\chi^2/ndf$\\[3pt] \hline\rule{0pt}{13pt}
$\overline{K}\rho$ & 1/3 & 2/3 &  0  & 7.0/3\\ 
$\overline{K}^*\pi$& 4/9 & 4/9 & 1/9 & 2.8/3\\ 
$K^-(\pi\pi)_{I=0}$  & 2/3 &  0  & 1/3 & 88.7/3\\ \hline\hline
\end{tabular}
\caption{\it 
The relative fraction for each $\overline{K}\pi\pi$ mode from isospin
symmetry for the possible intermediate states. The test of these 
fractions is given by the $\chi^2/ndf$ values for the hypothesis that
the given intermediate state dominates.} 
\label{isokpp}
\end{center}
\end{table}
\begin{table}
\begin{center}
\begin{tabular}{|c|c|c|c|c|}\hline\hline
Mode &$K^-\pi^+\pi^-~(10^{-3})$ 
& $\overline{K^0}\pi^-\pi^0~(10^{-3})$
&$K^-\pi^0\pi^0~(10^{-3})$ & $\chi^2/ndf$\\ [3pt] \hline\rule{0pt}{13pt}
$\overline{K}\rho$ & $0.83\pm0.35$
        & $2.42\pm0.57$
        &     0
        &  0.7/3\\     
$\overline{K}^*\pi$& $1.31\pm0.41$
        & $0.85\pm0.45$
        & $0.56\pm0.25$
        &  1.8/3\\ \hline\hline
\end{tabular}
\caption{\it Test of isospin symmetry for different intermediate states.
The separation of the branching ratios for $\overline{K}\rho$ 
and $\overline{K}^*\pi$ is based on the 
measurements of the $\overline{K}\rho$ fractions. 
The $\chi^2/ndf$ values represent the consistency with the expected
relative ratios given in Table~\ref{isokpp}.} 
\label{krho}
\end{center}
\end{table}

\subsection{The decay $\tau^-\to(\overline{K}\pi\pi\pi)^-\nu_\tau$}

The final states considered include $K^-\pi^+\pi^-\pi^0$, $K^-\pi^0\pi^0\pi^0$,
$K^0h^+h^-h^-$ and $\overline{K^0}\pi^-\pi^0\pi^0$,
in which the $K^-\eta$ contributions are excluded. Since 
the net strangeness for the $K^0h^+h^-h^-$ modes is not determined,
$\overline{K^0}\pi^+\pi^-\pi^-$ dominance is assumed. 
Isospin invariance for the $3\pi$ system
leads to only three symmetry classes~\cite{rouge,pais}:
two with $I_{3\pi}=1$, as shown
in Table~\ref{isok3pi}, and one with $I_{3\pi}=0$, expected to be
dominated by the $K^-\omega$ state. An axial-vector 
contribution from the decay $K^-_1(1270)\to K^-\omega\to K^-\pi^+\pi^-\pi^0$, 
amounting to $(0.59\pm0.19)\times10^{-3}$, was deduced in
Section 4.2 and is subtracted from the $K^-\pi^+\pi^-\pi^0$ mode.

The expected relative contributions from the $I_{3\pi}=1$ classes are 
compared to data in Table~\ref{isok3pi}.   
The present experimental precision does not allow  
definite conclusions. However, the different isospin 
configurations can be used to derive a better estimate of the total rate.
The two isospin possibilities lead to consistent values 
(see Table~\ref{isok3pi}) and 
the average using the Fermi statistical weights~\cite{pais}
yields (excluding the $K^-\omega$ mode)
\begin{equation}
B(\tau^-\to (\overline{K}\pi\pi\pi)^-\nu_\tau)=(0.76\pm0.44)\times10^{-3}~.
\end{equation}
\begin{table}
\begin{center}
\small
\begin{tabular}{|c|c|c|c|c|c|c|}\hline\hline
Class &$K^-\pi^+\pi^-\pi^0$ &$K^-\pi^0\pi^0\pi^0$&
$\overline{K^0}\pi^+\pi^-\pi^-$
&$\overline{K^0}\pi^-\pi^0\pi^0$ & $B(\overline{K}3\pi)~(10^{-3})$
&$\chi^2/ndf$\\ \hline 
$(300)$ & 2/15 & 1/5 & 8/15 & 2/15 & $0.65\pm0.39$ & 1.8/3\\ 
$(210)$ & 1/3  &  0  & 1/3  & 1/3  & $0.83\pm0.46$ & 3.0/3\\ \hline\hline
\end{tabular}
\caption{\it 
The relative fractions for each $\overline{K}\pi\pi\pi$ 
mode expected by isospin
invariance for $I_{3\pi}=1$. The three numbers $(n_1n_2n_3)$ label
Pais' isospin classes~\cite{pais}. 
The test of isospin symmetry is given by the $\chi^2/ndf$ values for
the hypothesis that the given intermediate state dominates.}
\label{isok3pi}
\end{center}
\end{table}

\subsection{The decay $\tau^-\to(K\overline{K}\pi)^-\nu_\tau$} 

For this mode, the decay dynamics are investigated using the
branching ratios for all the
relevant final states: $K^+K^-\pi^-$, $K^0_SK^-\pi^0$,
$K^0_LK^-\pi^0$, $K^0_SK^0_L\pi^-$ and $K^0_SK^0_S\pi^-$,
as shown in Table~\ref{brs1}. If the contribution of second-class currents
is ignored, the rates for $K^+K^-\pi^-$ and $K^0\overline{K^0}\pi^-$ 
are expected to be equal~\cite{rouge}, in agreement with
the measurement 
\begin{equation}
\frac{B(\tau^-\to K^0\overline{K^0}\pi^-\nu_\tau)}
{B(\tau^-\to K^+K^-\pi^-\nu_\tau)}=0.94\pm0.27~.
\end{equation}

\begin{figure}
\begin{center}
\vspace{-2.0cm}
\epsfxsize 10cm
\epsffile{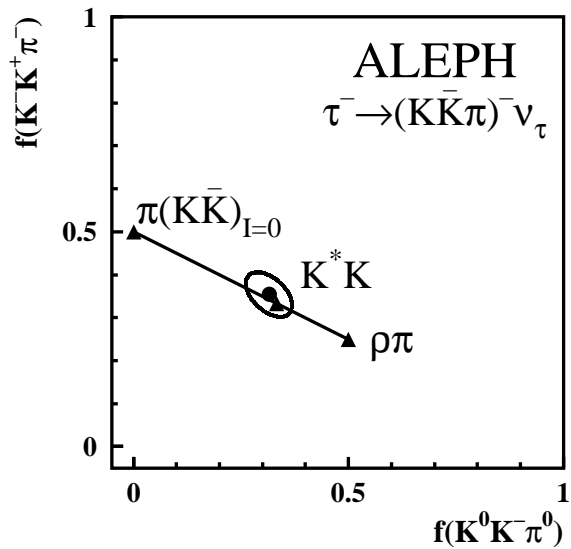}
 \end{center}
\vspace{-0.5cm}
\caption{\it The decay fractions
for $K^0K^-\pi^0\nu_\tau$ and
$K^-K^+\pi^-\nu_\tau$, given by a standard ellipse (39$\%$ probability),
are compared to the isospin constraint
on the $K\overline{K}\pi$ mode, shown by the straight line. The triangles
indicate the intermediate states dominated by $\pi(K\overline{K})_{I=0}$, 
$K^{*}K$ or $\rho\pi$.}
\label{fkkpi}
\end{figure}

The possible intermediate states appearing in
these decay modes are considered to be $\overline{K^*}K+K^*\overline{K}$ 
(denoted $K^*K$), $\rho\pi$ and $\pi(K\overline{K})_{I=0}$, 
corresponding to the fractions shown in Table~\ref{isokkpi}.
The test of isospin symmetry is consistent with the $K^*K$ 
intermediate state as illustrated in Fig.~\ref{fkkpi}, in agreement
with the conclusion from the analysis of the $K\pi$ mass spectra.
\begin{table}
\begin{center}
\begin{tabular}{|c|c|c|c|c|c|}\hline\hline
Mode &$K^-K^+\pi^-$ &$K^0_SK^0_L\pi^-$ &$K^0_SK^0_S\pi^-+K^0_LK^0_L\pi^-$
&$K^-K^0\pi^0$ & $\chi^2/ndf$\\ \hline 
$K^*K$    & 1/3 & 1/6 & 1/6 & 1/3 &  2.1/3\\ 
$\rho\pi$ & 1/4 & 1/4 &  0  & 1/2 & 16.0/3\\ 
$\pi(K\overline{K})_{I=0}$  & 1/2 &  0  & 1/2 &  0  & 99.0/3\\ \hline\hline
\end{tabular}
\caption{\it 
The relative fraction for each $K\overline{K}\pi$ mode in view of isospin
relations for three possible intermediate states. The test of these
relations is given by the $\chi^2/ndf$ values for the hypothesis
that the given intermediate state dominates.}
\label{isokkpi}
\end{center}
\end{table}

\subsection{The decay $\tau^-\to(K\overline{K}\pi\pi)^-\nu_\tau$} 

The limited experimental information (see Table~\ref{brs1}) on 
$K\overline{K}\pi\pi$ final states makes it impossible to study this channel
in any detail. Nevertheless, some conclusions can still be drawn. 
First, the rates for $\tau^-\to K^+K^-\pi^-\pi^0\nu_\tau$ and
$\tau^-\to K^0\overline{K^0}\pi^-\pi^0\nu_\tau$ are consistent within
their uncertainties, as expected by isospin symmetry and the absence of 
second-class currents~\cite{rouge}. Then, the relative 
fractions observed in the different modes can be compared with the 
expectations from the possible isospin configurations~\cite{rouge1},  
with a strong preference for the $(I_{K\overline{K}}=0,~I_{\pi\pi}=1)$ 
states and to a lesser extent $(I_{K\overline{K}}=1,~I_{\pi\pi}=1)$. 
The weighted average over the two non-zero modes yields an
estimate for the total rate
\begin{equation}
B(\tau^-\to(K\overline{K}\pi\pi)^-\nu_\tau)=(0.5\pm0.2)\times 10^{-3}~.
\end{equation}

\section{Vector and axial-vector separation}

The separation of the $\tau$ final states into vector and axial-vector
currents provides important information for QCD studies. Any deviation between
the inclusive sums of vector and axial-vector hadronic branching ratios
is necessarily a result of nonperturbative phenomena. In this section,
all the information obtained on the strange decay fractions 
concerning their $V/A$ character are summarized in order to determine
the corresponding inclusive rates. In addition, the separation of 
vector and axial-vector components is presented for 
the nonstrange $K\overline{K}\pi$ mode.

\subsection{$\tau^-\to (\overline{K}n\pi)^-\nu_\tau$}

The branching ratios for the strange sector of $\tau$ decays are 
globally reconsidered in
Table~\ref{rtausva}, with the respective current contributions 
unambiguously separated 
for the $K^-$, $(\overline{K}\pi)^-$, $K^-\eta$ and 
$(\overline{K}\pi\pi)^-$ modes.
In the $(\overline{K}3\pi)^-$ mode, the contribution expected 
from the axial-vector $K_1(1270)$ decay into $K^-\omega$ is subtracted.
The total $(\overline{K}4\pi)^-$ and $(\overline{K}5\pi)^-$ 
contributions are estimated on the basis of the 
branching ratios for the $5\pi$ and $6\pi$ modes~\cite{pdg98} with
Cabibbo and kinematic suppression, even though
the resulting total contribution is small in comparison to the precision
achieved in the measurement of the strange sector of $\tau$ decays.  
For the non-$K^-\omega$ part of $(\overline{K}3\pi)^-$, 
the $(\overline{K}4\pi)^-$ 
and the $(\overline{K}5\pi)^-$ contributions,
it is assumed that vector and axial-vector contribute equally with 
$100\%$ anti-correlated errors.
The corresponding branching ratios for vector and axial-vector final
states are
\begin{equation}
B(\tau^-\to V^-(S=-1)\nu_\tau)=(15.9~^{+1.6}_{-1.4})\times 10^{-3}
\end{equation}
and
\begin{equation}
B(\tau^-\to A^-(S=-1)\nu_\tau)=(12.8~^{+1.4}_{-1.7})\times 10^{-3}~.
\end{equation}
The total branching ratio for all strange $\tau$ decays
\begin{equation}
\label{totalbs}
B(\tau^-\to (V+A)^-(S=-1)\nu_\tau)=(28.7\pm1.2)\times 10^{-3}~
\end{equation}
is known with better precision. 

An estimate of the size of 
nonperturbative QCD contributions can be obtained from the ratio
\begin{equation}
\frac{B(\tau^-\to V^-(S=-1)\nu_\tau)-B(\tau^-\to A^-(S=-1)\nu_\tau)}
{B(\tau^-\to (V+A)^-(S=-1)\nu_\tau)}
=(11^{\,+10}_{\,-8})\%~,
\end{equation}
which is consistent with zero, but cannot rule out a significant 
nonperturbative component.
In comparison, the value $(1.7\pm1.0)\%$ was obtained 
in the nonstrange sector of $\tau$ decays~\cite{alphas}. 
\begin{table}
\begin{center}
\begin{tabular}{|c|c|c|c|}\hline\hline
Mode        & $B_V~(10^{-3})$ & $B_A~(10^{-3})$ 
            & $B_{V+A}~(10^{-3})$ \\ \hline
$K^-$  & $-$               & $ 6.96\pm0.29$ 
            & $ 6.96\pm0.29$ \\ 
$(\overline{K}\pi)^-$  & $13.60\pm0.62$    & $-$          
            & $13.60\pm0.62$ \\
$(\overline{K}2\pi)^-$ & 
            $ 1.39^{\,+1.30}_{\,-1.01}$ &$4.58^{\,+1.23}_{\,-1.49}$
            & $ 5.97\pm0.73$ \\
$K^-_1(1270)\to K^-\omega$& $-$ & $ 0.67\pm0.21$ 
            & $ 0.67\pm0.21$ \\
$K^-\eta$   & $ 0.29^{\,+0.15}_{\,-0.14}$ & $-$
            & $ 0.29^{\,+0.15}_{\,-0.14}$\\
$(\overline{K}3\pi)^-$ & $ 0.38\pm0.53$    & $ 0.38\pm0.53$
            & $ 0.76\pm0.44$ \\
$(\overline{K}4\pi)^-$ & $ 0.17\pm0.37$    & $ 0.17\pm0.37$
            & $ 0.34\pm0.34$ \\
$(\overline{K}5\pi)^-$ & $ 0.03\pm0.10$    & $ 0.03\pm0.10$
            & $ 0.06\pm0.06$ \\ \hline
Sum & $15.86^{\,+1.60}_{\,-1.37}$ & $12.79^{\,+1.43}_{\,-1.66}$
    & $28.65\pm1.17$ \\ \hline\hline
\end{tabular}
\caption{\it Branching ratios for vector and axial 
vector current contributions to the strange sector of $\tau$ decays.
The branching fractions for the 
$\overline{K}4\pi$ and $\overline{K}5\pi$ 
modes are obtained from the measured branching 
ratios for the $5\pi$ and $6\pi$ in Ref.~\cite{pdg98} introducing
the Cabibbo suppression and kinematic factors. Vector and axial
vector currents in the $\overline{K}3\pi$, $\overline{K}4\pi$ and
$\overline{K}5\pi$ are assumed to contribute equally.}
\label{rtausva}
\end{center}
\end{table}

\subsection{$\tau^-\to(K\overline{K}\pi)^-\nu_\tau$}

In the analysis of the nonstrange spectral 
functions\cite{vect,alphas}, fractions of $(50\pm50)\%$ vector and
axial-vector currents were assumed for lack of a better knowledge 
in $\tau^-\to(K\overline{K}\pi)^-\nu_\tau$ decays. 
Due to the anti-correlation between vector and axial-vector, this 
uncertainty was important for the difference $V-A$. 
\vs
In this analysis, two methods are considered in order to 
separate the vector and axial-vector 
contributions in the $K\overline{K}\pi$ mode. The first method based on
the G-parity~\cite{rouge} applies best to the cases of 
either a pure vector or a pure axial-vector state. 
The second method is to compare the inclusive $V+A$ $K\overline{K}\pi$ 
spectral function of $\tau$ decays with the 
corresponding isovector cross section measurements 
in $e^+e^-$ annihilation, providing the $V$ part by means of CVC.

The G-parity of the $K\overline{K}\pi$ system is 
determined by the $K\overline{K}$ system with
$G_{K\overline{K}}=(-1)^{I_{K\overline{K}}}(-1)^{l_{K\overline{K}}}$. 
Since the $K^0K^-\pi^0$ mode can only have 
$I_{K\overline{K}}=1$, the corresponding decay width amounts to
half of the total decay width for $I_{K\overline{K}}=1$. 
The isospin configuration $I_{K\overline{K}}=0$ contributes to both the 
$K^0\overline{K^0}\pi^-$ and $K^+K^-\pi^-$ modes. 
In addition, the $l_{K\overline{K}}$ values are always odd for 
$K^0_SK^0_L$ and even for $K^0_SK^0_S$ or $K^0_LK^0_L$. The two following
ratios are therefore defined~\cite{rouge}:
\begin{equation}
r_0=\frac{\Gamma_{K^0_SK^0_S\pi^-}+\Gamma_{K^0_LK^0_L\pi^-}}
{\Gamma_{K^0_SK^0_L\pi^-}}=
\frac{2\Gamma^0_A+\Gamma^1_V}{2\Gamma^0_V+\Gamma^1_A}~,
\end{equation}
and
\begin{equation}
r=\frac{\Gamma_{K^0\overline{K^0}\pi^-}+\Gamma_{K^+K^-\pi^-}
-\Gamma_{K^0K^-\pi^0}}{\Gamma_{K^0K^-\pi^0}}
=2\frac{\Gamma^0_A+\Gamma^0_V}{\Gamma^1_A+\Gamma^1_V}~,
\end{equation}
where $\Gamma^0_{V,A}$ and $\Gamma^1_{V,A}$ are the decay widths for the 
$I_{K\overline{K}}=0$ and $I_{K\overline{K}}=1$ states, respectively. 
For pure vector, one has $1/r=r_0$, 
whereas $r=r_0$ for pure axial vector. 
The branching ratio measurements from ALEPH give 
\begin{equation}
r_0=0.52^{\,+0.32}_{\,-0.25}
\end{equation}
and
\begin{equation}
r=1.19^{\,+0.65}_{\,-0.48}~.
\end{equation}
Consequently, the probabilities for the $K\overline{K}\pi$ mode being 
a pure axial vector and a pure vector are computed to be about $25\%$ and 
$50\%$, respectively. It is therefore not possible to draw a firm conclusion
from this study. This is not too surprising as the $K\overline{K}\pi$ mode is 
observed to be dominated by $K^*K$, in which case the previous test becomes
degenerate. 
\begin{figure}[t]
\begin{center}
\vspace{-2.0cm}
\epsfxsize 10cm
\epsffile{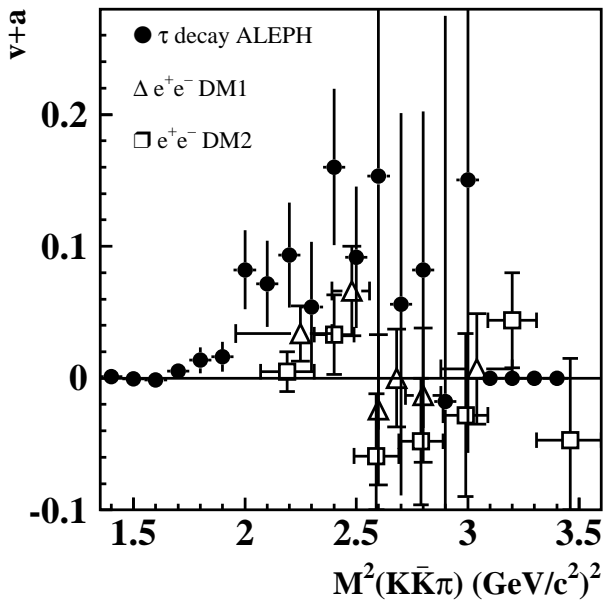}
\end{center}
\vspace{-0.5cm}
\caption{\it The $V+A$ spectral function using
the $K\overline{K}\pi$ data from $\tau$ decays
compared to those derived from the isovector $e^+e^-$ cross sections. 
The $e^+e^-$  data are taken from DM1~\cite{dm1} and DM2~\cite{dm2}.}
\label{spectkkp}
\end{figure}

Another possibility is the comparison between the measured spectral function
in $\tau$ decay and the measured isovector cross sections in 
$e^+e^-\to K\overline{K}\pi$. The application of CVC makes it 
possible to link the $\tau$ decays into $K\overline{K}\pi$ final states 
with $e^+e^-$ annihilation into
isovector states $K^0_SK^\pm\pi^\mp$ and $K^\pm K^\mp\pi^0$, 
via the relation
\begin{equation}
\sigma^{I=1}_{e^+e^-\to K\overline{K}\pi}
=\frac{4\pi\alpha^2}{s}v_{1,K\overline{K}\pi\nu_\tau}~,
\label{cross}
\end{equation}
where $\sigma^{I=1}_{e^+e^-\to K\overline{K}\pi}$ is the 
isovector cross section, $\alpha$ is the electromagnetic 
fine structure constant, and $v_{1,K\overline{K}\pi\nu_\tau}$
is the corresponding vector $\tau$ spectral function.
There is only isovector in $\tau$ decay,  
while the isospin for the $K\overline{K}\pi$ 
system in $e^+e^-$ annihilation 
can be either 0 or 1. By studying the Dalitz plot,
the DM1~\cite{dm1} and DM2~\cite{dm2} experiments 
observed $K^*K$ dominance and concluded
that the reaction $e^+e^-\to K\overline{K}\pi$ is dominated by the 
isoscalar (resonant) amplitude, whereas a small 
isovector component is extracted from the interference of both amplitudes.

The measured invariant mass-squared 
distribution of the $K\overline{K}\pi$ system in $\tau$ decay 
(Fig.~\ref{mkkp}) is transformed into the $V+A$ spectral 
function (according to Eq.~(\ref{spect}) in the next section).
The cross section measurements for $e^+e^-\to K\overline{K}\pi$
~\cite{dm1,dm2} are also converted into a spectral function
according to Eq.~(\ref{cross}). Since the DM1 and DM2 experiments
only provide studies of $K^0_SK^\pm\pi^\mp$, the full $K\overline{K}\pi$ 
spectrum is obtained by scaling up the $K^0_SK^\pm\pi^\mp$
contribution by a factor of three. Figure~\ref{spectkkp}
shows the comparison between $\tau$ decay and 
$e^+e^-$ annihilation. A fairly large excess is observed
below 2.5 (GeV/$c^2$)$^2$ compared to the $I=1$ $e^+e^-$ data.
The measured cross sections
of $e^+e^-\to K\overline{K}\pi$ are translated by CVC
into the expected vector branching ratio 
$B_V(\tau^-\to (K\overline{K}\pi)^-\nu_\tau)$, yielding 
$(0.67\pm0.27)\times 10^{-3}$ for the DM1 measurement and
$-(0.12\pm0.26)\times 10^{-3}$ for the DM2 measurement, which are averaged 
to $(0.26\pm0.39)\times 10^{-3}$. This value is much
smaller than the measured branching ratio 
$B(\tau^-\to (K\overline{K}\pi)^-\nu_\tau)=(4.60\pm0.50)\times10^{-3}$, 
yielding a dominant axial-vector fraction of $(94^{\,+6}_{\,-8})\%$.

One can combine all the information sensitive to 
the V and A fractions: relative rates of different 
$K\overline{K}\pi$ modes in $\tau$ decay and comparison
of total rates in $\tau$ decays and $I=1$ $e^+e^-$ annihilation. It is found
that the $I_{K\overline{K}}=1$ component in the dominant axial-vector part is
$(75\pm9)\%$, while in the much smaller vector part it amounts to      
$(43\pm28)\%$. These fractions agree with the expected value of 
2/3 for $K^*K$ dominance, established for the axial-vector
current in this paper, and for the vector current in $e^+e^-$ 
annihilation~\cite{dm1,dm2}.

A natural candidate to explain the dynamics of the $K\overline{K}\pi$
mode is the decay $a_1 \to K^*K$. In fact the observed mass spectrum
agrees well with the Monte Carlo prediction 
based on this model (Fig.~\ref{mkkp}), 
showing a sharp rise at the $K^*K$ threshold. Under the assumption
of $a_1$ dominance and using the branching ratio 
$B(\tau \to a_1 \nu_\tau)$~\cite{alephbr_had}, one gets the branching
fraction $B(a_1 \to K^*K) = (2.6 \pm 0.3)\%$. This value is in good
agreement with the preliminary result from a partial wave 
analysis~\cite{cleopwa1} in the $\tau \to \pi \pi^0 \pi^0 \nu_\tau$
channel, including the opening of the $K^*K$ decay channel in the $a_1$
total width, yielding $B(a_1 \to K^*K) = (3.3 \pm 0.5 \pm 0.1)\%$.
This observation provides an additional and independent argument
for axial-vector dominance in the $K\overline{K}\pi$ mode.

The observed axial-vector dominance is in qualitative agreement with
the models developed in Ref.~\cite{fink,dms} and in contradiction
with those proposed in Ref.~\cite{jjg,bal}.  

\section{The total strange spectral function}

Theoretically, the hadronic $\tau$ decay width 
can be formulated in terms of the spectral functions~\cite{yst}
$v_1(s)$, $a_1(s)$ and $a_0(s)$ for the nonstrange part, 
and $v_1^S(s)$, $a_1^S(s)$, $v_0^S(s)$ and $a_0^S(s)$ for the strange part, 
where $s$ is the hadronic mass-squared. 
The notations $v$ and $a$ stand for vector and axial-vector, 
while the subscript refers to the spin $J$ of the hadronic system. 
These spectral functions can be
experimentally determined by measuring the invariant mass spectra of
given hadronic modes and normalizing them to their respective branching 
ratios. The nonstrange spectral functions ($v_1/a_1$ and $a_0$) 
are defined and measured in~\cite{vect,alphas}, and the strange 
spectral functions read
\begin{eqnarray}
v_1^S(s)/a_1^S(s)&\equiv&\frac{M^2_\tau}{6|V_{us}|^2S_{EW}} 
\frac{B(\tau^-\to V^-/A^-(S=-1,J=1)\nu_\tau)}
{B(\tau^-\to e^-\bar{\nu}_e\nu_\tau)}\nonumber \\
& &\times\frac{1}{N_{V/A}}\frac{dN_{V/A}}{ds}
\left[\left(1-\frac{s}{M^2_\tau}\right)^2
\left(1+\frac{2s}{M^2_\tau}\right)\right]^{-1}
\label{spect}
\end{eqnarray}
and
\begin{eqnarray}
v_0^S(s)/a_0^S(s)&\equiv&\frac{M^2_\tau}{6|V_{us}|^2S_{EW}} 
\frac{B(\tau^-\to V^-/A^-(S=-1,J=0)\nu_\tau)}
{B(\tau^-\to e^-\bar{\nu}_e\nu_\tau)}\nonumber \\
& &\times\frac{1}{N_{V/A}}\frac{dN_{V/A}}{ds}
\left(1-\frac{s}{M^2_\tau}\right)^{-2}~,
\end{eqnarray}
where $|V_{us}|=0.2218\pm0.0016$~\cite{pdg98},
$M_\tau=(1776.9^{\,+0.31}_{\,-0.27})$ MeV/$c^2$ 
is taken from the BES measurement~\cite{bes},
$S_{EW}=1.0194\pm0.0040$ accounts for the electroweak radiative 
corrections~\cite{ew}, and $dN_{V/A}/N_{V/A}ds$
denotes the normalized distribution of the invariant mass-squared of the
corresponding vector/axial-vector decay channels $V/A$.
The leptonic branching ratio is taken to be 
$B(\tau^-\to e^-\bar{\nu}_e\nu_\tau)=(17.794\pm0.045)\%$, where the 
precision has been improved by applying lepton universality as 
described in Ref.~\cite{alphas}. The dominant contribution
to $a_0^S(s)$ is expected to be provided by the single kaon pole
with $a_{0,K}^S(s)=4\pi^2f^2_K\delta(s-M^2_K)$.

Because the spectral functions measure the transition
probability to create hadrons with a mass $\sqrt{s}$ from the QCD vacuum,
they are the natural input to QCD studies, allowing tests
to be performed at a running mass scale less than or equal to $M_\tau$. Their 
nonstrange vector part can be directly related to the low energy 
$e^+e^-$ annihilation cross section using CVC. In addition,
the strange spectral function carries  
information on the strange quark mass.     

\begin{figure}[t]
\begin{center}
\vspace{-2.0cm}
\epsfxsize 14cm
\epsffile{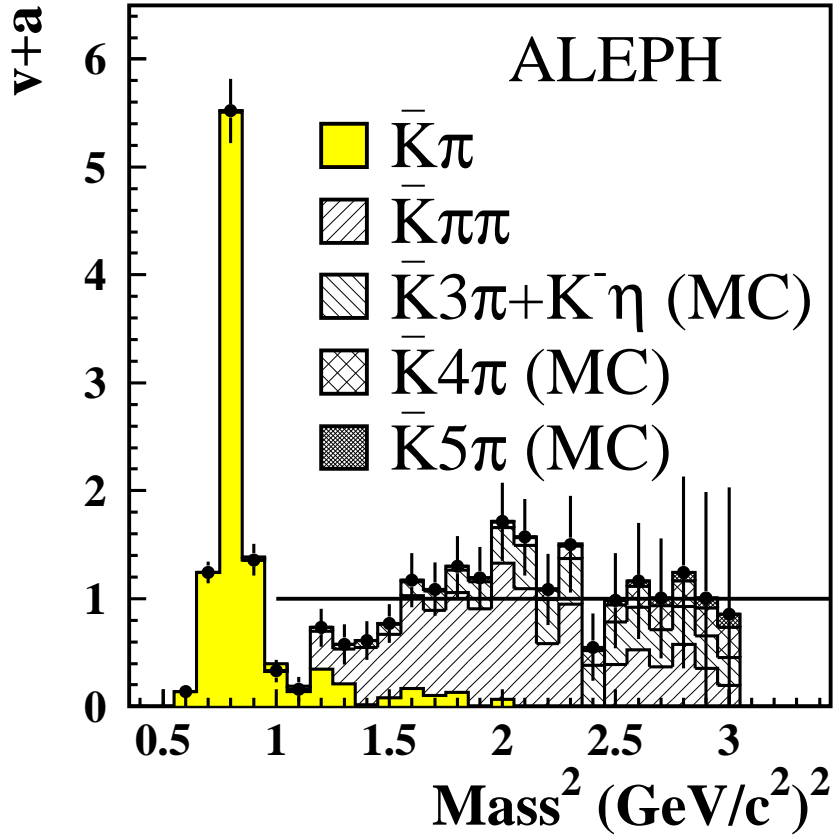}
\end{center}
\vspace{-0.5cm}
\caption{\it Total hadronic vector and axial-vector spectral
function (dots with error bars) from $\tau$ decays into strange final states
with its different contributions indicated. The errors include systematic
uncertainties. The kaon pole is not shown in this plot. 
The shapes for the $K^-\eta$ and $\overline{K}n\pi(n\geq3)$ modes are obtained 
from the Monte Carlo simulation. 
The parton model prediction is given by the solid straight line.}
\label{fig_specfun}
\end{figure}

The total strange spectral function 
is shown in Fig.~\ref{fig_specfun}. 
The spectra for the $\overline{K}\pi$ and 
$\overline{K}\pi\pi$ modes are obtained from the
corrected mass spectra, taking into account the acceptance and bin migration
corrections. For the other modes a phase space 
generator is used to simulate the corresponding mass distributions.
The contributions from the respective 
channels are normalized to the corresponding branching ratios
according to Eq.~(\ref{spect}).

The $K^*(892)^-$ resonance stands 
out clearly on the low mass side, while within a large uncertainty 
(dominated by background subtraction in the $(\overline{K}\pi\pi)^-$ modes)
the higher energy part is consistent with the  parton model expectation.

The total strange spectral function is used in the following section,
in order to construct spectral moments needed for the QCD analysis.

\section{The strange quark mass}

A main goal of the present analysis is the determination of the 
strange quark mass \ms.
The method adopted follows the line of the recent ALEPH \asm\
determination from nonstrange hadronic $\tau$ decays~\cite{alphas} and
is based on a simultaneous fit of QCD parametrizations, including
perturbative and nonperturbative components, to measured observables.
A more detailed account of this method and of the phenomenological
context can be found in Ref.~\cite{cdh}.

\subsection{The strange hadronic decay ratio \RtS}

As previously demonstrated in Ref.~\cite{davchen}, the inclusive 
$\tau$ decay ratio into strange hadronic final states,
\beq
     R_{\tau,S} =
\frac{\Gamma(\tau^-\rightarrow{\rm hadrons}_{S=-1}^-\,\nu_\tau)}
                   {\Gamma(\tau^-\rightarrow e^-\,\bar{\nu}_e\nu_\tau)}~,
\eeq
can be used due to its precise theoretical prediction~\cite{bnp,chetkwiat}
in the framework of the Operator Product Expansion (OPE)~\cite{svz} to 
determine \mss\ at the scale $s=M_\tau^2$. Since then it was 
shown~\cite{maltman} that the perturbative expansion used for the massive 
term in Ref.~\cite{chetkwiat} was incorrect. After correction the series
shows a problematic convergence behaviour~\cite{maltman,pichprad,kuhnchet}.
\vs
Following Ref.~\cite{bnp} the theoretical prediction for the inclusive 
vector plus axial-vector hadronic decay rate is given by
\beqn
\label{eq_rtau}
   R_\tau(M_\tau^2) &\;=\;&
     12\pi S_{\rm EW}\intl_0^{M_\tau^2}
     \frac{ds}{M_\tau^2}\left(1-\frac{s}{M_\tau^2}\right)^{\!\!2}\nonumber\\
     &&\times\left[\left(1+2\frac{s}{M_\tau^2}\right){\rm Im}\Pi^{(1+0)}(s)
      \,-\,2\frac{s}{M_\tau^2}{\rm Im}\Pi^{(0)}(s)\right]~,
\eeqn
with the two-point correlation functions
$\Pi^{(J)}=|V_{ud}|^2\Pi_{ud,V+A}^{(J)}+|V_{us}|^2\Pi_{us,V+A}^{(J)}$
for the hadronic final state of the spin $J$. The choice of the particular
spin combination for the correlators taken in Eq.~(\ref{eq_rtau}) is
justified in Ref.~\cite{bnp}. Equation~(\ref{eq_rtau}) can be decomposed as
\beq
\label{eq_delta}
   R_{\tau,(S)} \;=\;
     3|V|^2S_{\rm EW}\left(1 + \delta^{(0)} + 
     \delta^{(2-\rm mass)} + 
     \hm\hm\sum_{D=4,6,\dots}\hm\hm\hm\hm\delta^{(D)}+
     \delta^\prime_{\rm EW}\right)~,
\eeq
where $V=V_{ud}$ ($V=V_{us}$) for the nonstrange (strange) case
is the corresponding CKM matrix element. The residual non-logarithmic 
electroweak correction $\delta^\prime_{\rm EW}\simeq0.0010$~\cite{braaten}
is neglected in the following. Throughout this analysis, all QCD observables
are expressed in the $\rm{\overline{MS}}$ renormalization scheme.  

\subsubsection{Perturbative contributions}

The $\delta^{(0)}$ term in Eq.~(\ref{eq_delta})
is the perturbative part of mass dimension $D=0$,
known to third order~\cite{3loop} in the expansion
with $a_s(M_\tau^2)=\alpha_s(M_\tau^2)/\pi$, and studied in detail in
Ref.~\cite{alphas}.
\vs
Next, the $ \delta^{(2-\rm mass)}$ term is the mass
contribution of dimension $D=2$,  in practice only important for the
strange quark mass and the subject of the present study. The
fixed-order perturbation theory (FOPT) gives~\cite{chetkwiat,maltman}
\beqn
\label{eq_delta2}
   \delta^{(2-\rm mass)}_S &=&-\,8\frac{m^2_s(M_\tau^2)}{M_\tau^2}
     \Bigg[ 1 + \frac{16}{3}a_s(M_\tau^2) + 46.00\,a_s^2(M_\tau^2)\nonumber \\
& &+ \left(283.6 + \frac{3}{4}\,x_3^{(1+0)}\right)a_s^3(M_\tau^2)\Bigg]~.
\eeqn
The third order coefficient $x_3^{(1+0)}$ occurs in the expansion of 
the massive $J=1+0$ correlator~\cite{chetkwiat,chet,chetkuhn},
while the $J=0$ correlator is known up to third order.
Assuming the perturbative expansions to behave like a
geometric series, one can estimate the unknown coefficient to be
$x_3^{(1+0)}\approx x_{2}^{(1+0)}(x_{2}^{(1+0)}/x_{1}^{(1+0)})
=165\pm330$, using as error twice the estimated 
contribution. Setting 
$\alpha_s(M_\tau^2)=0.334$~\cite{alphas}, Eq.~(\ref{eq_delta2}) 
becomes
\beq
\label{eq_convfopt}
   \delta^{(2-\rm mass)}_S \;=\; 
     -\,8\frac{m^2_s(M_\tau^2)}{M_\tau^2}
     \left[1 + 0.57 + 0.52 + (0.49\pm0.29) \right]~,
\eeq
which converges badly.

A so-called {\it contour-improved} FOPT~\cite{pert,pivo} analysis 
(\FOPTCI) for the dimension $D=2$ contribution 
has been presented in Refs.~\cite{pichprad,kuhnchet}. 
It consists of a direct numerical evaluation of the contour integral 
derived from Eq.~(\ref{eq_rtau}), using the solution of the 
renormalization group equation (RGE) to four loops~\cite{rit,ritmass} as input 
for the running $\alpha_s(s)$ and $m_s(s)$. This provides a resummation of 
all known higher order logarithmic integrals and was observed to improve 
the convergence of the massless perturbative series~\cite{alphas,pert}. 
With the above value for $\alpha_s$,
the contour-improved evaluation of the perturbative series reads
\beq
\label{eq_convfoptci}
   \delta^{(2-\rm mass)}_{S,CI} \;=\; 
     -\,8\frac{m^2_s(M_\tau^2)}{M_\tau^2}
     \left[0.97 + 0.49 + 0.38 + (0.34\pm0.04) \right]~,
\eeq
with somewhat improved convergence compared to 
FOPT~(\ref{eq_convfopt}). Therefore, the results of the following analysis are 
based on the \FOPTCI\ approach.
\vs
Independently of whether FOPT or \FOPTCI\ is used, the origin of the
convergence problem is found in the $J=0$ component as defined 
in Eq.~(\ref{eq_rtau})
\beq
\label{eq_cfoptci0}
   \delta^{(2-\rm mass)}_S(J=0) \;=\; 
     -\,8\frac{m^2_s(M_\tau^2)}{M_\tau^2}
     \left[0.41 + 0.32 + 0.32 + 0.37 \right]~,
\eeq
while the $J=1+0$ part converges very well,
\beq
\label{eq_cfoptci10}
   \delta^{(2-\rm mass)}_S(J=1+0) \;=\; 
     -\,8\frac{m^2_s(M_\tau^2)}{M_\tau^2}
     \left[0.56 + 0.16 + 0.05 - (0.04\pm0.04) \right]~.
\eeq

Following these observations, two methods are considered in the following
in order to determine $m_s(M_\tau^2)$:
\begin{itemize}
\item in the {\it inclusive method}, the inclusive strange
hadronic rate is considered and both $J=1+0$ and $J=0$ are included
with their respective convergence behaviour taken into account
in the theoretical uncertainties.

\item the {\it `1+0' method} singles out the 
well-behaved $J=1+0$ part by subtracting the experimentally
determined $J=0$ longitudinal component from data. The measurement is then
less inclusive and the sensitivity to $m_s$ is significantly
reduced; however, the $\delta_2$ perturbative expansion is under
control and the corresponding theoretical uncertainty is reduced.
\end{itemize}

Since the $J=0$ expansion is problematic, the results 
are given with the `1+0' method, while the inclusive approach is
used as an insight into the handling of Eq.~(\ref{eq_cfoptci0}).

\subsubsection{Nonperturbative contributions}
The last term in Eq.~(\ref{eq_delta}) represents the 
nonperturbative contribution which, using the OPE, can be
written as a sum of powers of $M_\tau^D$:
\beq
\label{eq_ope}
    \delta^{(D)} \;=\;
       \hm\hm\hm\sum_{{\rm dim}{\cal O}=D}\hm\hm\hm C(s,\mu)
            \frac{\langle{\cal O}(\mu)\rangle}
                 {(-M_\tau^2)^{D/2}}~,
\eeq
where the parameter $\mu$ separates the 
long-distance nonperturbative ef\/fects, absorbed into
the vacuum expectation elements $\langle{\cal O}(\mu)\rangle$,
from the short-distance ef\/fects which are included in the
Wilson coef\/f\/icients $C(s,\mu)$~\cite{wilson}.

 The dimension $D=4$  operators have dynamical 
contributions from the gluon condensate \GG, the quark 
condensates $m_u\langle{\bar u}u\rangle$, 
$m_d\langle{\bar d}d\rangle$ and $m_s\langle{\bar s}s\rangle$,
and the running quark masses to the fourth power. 
The contributions from dimension $D=6$ and $D=8$ operators are rather 
complex and they are taken into account by introducing ef\/fective scale 
independent operators $\langle{\cal O}_{6}\rangle$ and
$\langle{\cal O}_{8}\rangle$ that are f\/itted to the data.

\subsection{Evidence for the effect of a massive strange quark}

From the result found from this analysis on the branching ratio
for $\tau$ decays into all strange hadronic final states~(\ref{totalbs}),
one obtains
\beq
\label{rtausexp}
   R_{\tau,S} = 0.1610 \pm 0.0066~,
\eeq
using the value for $B(\tau^- \rightarrow e^- \bar{\nu}_e \nu_\tau)$
given in Section 6.

The result~(\ref{rtausexp}) can be readily compared with the
QCD prediction for a massless strange quark neglecting the nonperturbative
contributions,
\beq
\label{rtausth0}
   R^{(0)}_{\tau,S} = 0.1809 \pm 0.0036~, 
\eeq
where the quoted error mainly reflects the 
uncertainties on $\alpha_s(M^2_\tau)$
and $V_{us}$. The two values differ by $2.7 \sigma$, in
the direction predicted for a non-zero $m_s$ value.

This is evidence for the effect of a massive strange quark in
a single observable which is essentially predicted by perturbative QCD,
providing the basis for the determination of the running mass $m_s$ at 
the scale of the $\tau$ mass.

\subsection{Spectral moments}

As proposed in Ref.~\cite{pichledib} and successfully applied in
several \asm\ analyses~\cite{alphas,aleph_as,cleo_as,opal_as},
the \sfs\ are used to construct the moments
\beq
\label{eq_moments}
   R_{\tau,(S)}^{kl} \;\equiv\; 
       \intl_0^{M_\tau^2} ds\,\left(1-\frac{s}{M_\tau^2}\right)^{\!\!k}
                              \left(\frac{s}{M_\tau^2}\right)^{\!\!l}
       \frac{dR_{\tau,(S)}}{ds}~,
\eeq
with $R_{\tau,(S)}^{00}=R_{\tau,(S)}$ for the nonstrange and the
strange cases, respectively. 
The theoretical prediction for the moments~(\ref{eq_moments})
follows the line described above, leading to expressions
similar to Eq.~(\ref{eq_delta}) with nonperturbative 
contributions $\delta^{kl(D)}$~\cite{cdh}.
Fitting the $\tau$ decay rate and the spectral moments allows $m_s(M_\tau^2)$
and the nonperturbative operators of dimension $D=6$ and $D=8$ 
to be simultaneously obtained.

In order to reduce the theoretical uncertainties,
the difference between nonstrange and strange spectral 
moments, properly normalized with their respective CKM matrix elements,
is considered:
\beq
\label{eq_dmoments}
   \Delta_\tau^{kl} \;\equiv\;
     \frac{1}{|V_{ud}|^2}R_{\tau,S=0}^{kl} - 
     \frac{1}{|V_{us}|^2}R_{\tau,S=-1}^{kl}~,
\eeq
for which the massless perturbative contribution vanishes so that the
theoretical prediction now reads (setting $m_u=m_d=0$)
\beq
\label{eq_Deltamom}
   \Delta_\tau^{kl} \;=\;
     3S_{\rm EW}\left(
     -\,\delta_{S}^{kl(2-\rm mass)} + 
     \hm\hm\sum_{D=4,6,\dots}\hm\hm\hm\hm\tilde{\delta}^{kl(D)}\right)~.
\eeq
For the dimension $D=4$ nonperturbative contribution, 
the flavour independent gluon condensate disappears in the difference,
leaving only the quark condensate terms. The fitted 
$\tilde{\delta}^{kl(6,8)}$ contain the higher-order nonperturbative 
contributions to the difference of nonstrange and strange moments.
Since the nonstrange nonperturbative contributions to the inclusive 
$V+A$ hadronic decay ratio have been found to be very small~\cite{alphas}, 
one has $\tilde{\delta}^{kl(4,6,8)}\approx-\delta_{S}^{kl(4,6,8)}$.

In this analysis, the nonstrange spectral function obtained from
Ref.~\cite{alphas} has been appropriately scaled making use of  
the updated value of $R_{\tau,S}$\rlap.\footnote{$R_{\tau,S=0}$ is obtained
from the difference $R_\tau-R_{\tau,S}$, in which $R_\tau$ is directly 
derived from the leptonic branching ratio quoted in Section~6.}
The experimental results of the spectral moments~(\ref{eq_dmoments})
and their correlations are given in Table~\ref{tab_dmoments}. For the
CKM matrix elements the values $|V_{ud}|=0.9751\pm0.0004$ and 
$|V_{us}|=0.2218\pm0.0016$~\cite{pdg98} are used, while the errors are
included in the theoretical uncertainties discussed in the next section.
In the inclusive $V+A$ nonstrange case the correlations between the hadronic
decay ratio, obtained essentially using universality from the 
leptonic branching ratios and the $\tau$ lifetime, and the 
spectral moments are negligible (see Ref.~\cite{alphas}). On the contrary, 
all $(k,l)$ moments used here suffer from large correlations
due to the common input from the strange spectral function. This reduces
the independent information used in the fit and thus generates
strong correlations between the adjusted parameters.
\begin{table}[t]
\centerline{
\begin{tabular}{|cc|} 
\hline\hline 
 $(k,l)$& $\Delta_\tau^{kl}$ \\
\hline
  (0,0) & $0.394 \pm 0.137$ \\
  (1,0) & $0.383 \pm 0.078$ \\
  (2,0) & $0.373 \pm 0.054$ \\
  (1,1) & $0.010 \pm 0.029$ \\
  (1,2) & $0.006 \pm 0.015$ \\
\hline\hline
\end{tabular}
\hspace{1cm}
\begin{tabular}{|cccccc|} \hline\hline
$(k,l)$ & (0,0) & (1,0) & (2,0) & (1,1) & (1,2)  \\
\hline
(0,0)   &  1    &  0.94 & 0.83  & 0.98  & 0.91   \\
(1,0)   & --    &  1    & 0.97  & 0.87  & 0.72   \\
(2,0)   & --    & --    &  1    & 0.72  & 0.53   \\
(1,1)   & --    & --    & --    &  1    & 0.95   \\
(1,2)   & --    & --    & --    & --    & 1      \\
\hline\hline
\end{tabular}
}
\caption[.]{\label{tab_dmoments}\it
            Measured spectral moments $\Delta_\tau^{kl}$
            (left table) and their experimental correlations
            (right table).}
\end{table}

\begin{figure}
\begin{center}
\vspace{-2.0cm}
\epsfxsize 10cm
\epsffile{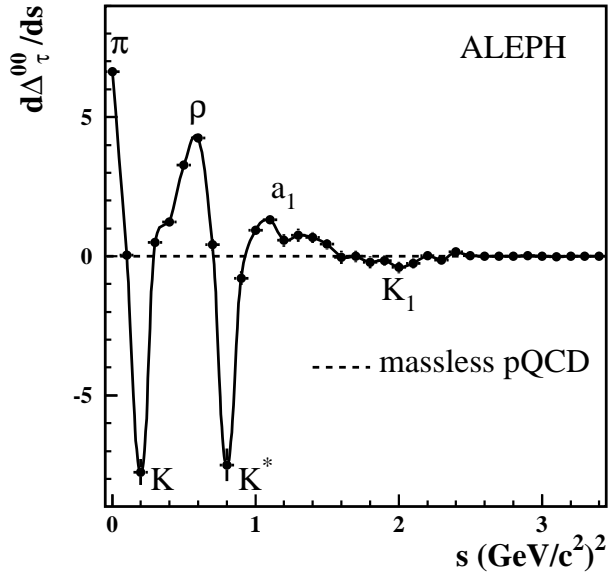}
\end{center}
\vspace{-0.5cm}
\caption{\it Integrand of Eq.~(\ref{eq_dmoments}) for (k=0, l=0),
$i.e.$, difference of the Cabibbo corrected nonstrange and strange 
invariant mass spectra. The contribution from massless perturbative 
QCD vanishes in the difference.}
\label{diffmoms}
\end{figure}

Figure~\ref{diffmoms} shows the weighted integrand of the lowest 
moment $\Delta^{00}_\tau$ (see Eq.~(\ref{eq_moments}),(\ref{eq_dmoments})) 
from the ALEPH data, as 
a function of the invariant mass-squared, and for which 
the expectation from massless perturbative QCD vanishes.

\subsection{Separation of $J=0,1$ components}

Although the $J=0$ component has not been
thoroughly investigated in the data, it is clear that it is
dominated by the single $K$ pole. Other contributions can be
identified: {\it (i)} off-shell vector $K^*$ resonances can generate
a $0^+$ component~\cite{finkescalar}, however too small to have 
a significant effect in this analysis, and {\it (ii)} production of the
scalar $K^*_0(1430)$ resonance may occur as discussed below.

The $K^*_0(1430)$ state decays almost exclusively into $K \pi$ with a
branching ratio of $(93 \pm 10) \%$~\cite{pdg98} and a fit to the
$K^- \pi^0$ mass distribution as discussed in Section 4.1 provides
a branching ratio 
$B(\tau \rightarrow K^*_0(1430)\nu_\tau)~=~ (0.0 \pm 2.5) \times 10^{-4}$.
It is therefore clear that $0^+$ contributions are strongly suppressed
($< 1.6 \times 10^{-2}$) compared to the dominant $1^-$ production. This
is expected from chiral symmetry breaking, with contributions reduced
by $\sim (m_K / M_\tau)^4$ as discussed in Ref.~\cite{finkescalar}, not
including the $\tau$ kinematic factor which disfavours higher mass
states.

No direct estimate of extra $0^-$ components beyond the single $K$ can
be made from the present data in the $\overline K \pi \pi$ modes. This is
due to the lack of statistics and to the background subtraction, 
especially in the high mass region where some evidence for 
a broad pseudoscalar state at $1460~{\rm MeV}/c^2$ exists~\cite{pdg98}.
A contribution equal to the $0^+$ one (actually an upper limit) 
is assumed in this case.

The evaluation of the $J=0$ contributions to $R_{\tau,S}$ obtained from
integration of Eq.~(\ref{eq_rtau}) using $f_K$ from Eq.~(\ref{fkdecont}) gives 
\beqn
\label{eq_spin0K}
         R_{\tau,S}^{(0)}(K)~&=&~ -0.00615 \pm 0.00026~, \\
\label{eq_spin0rest}
         R_{\tau,S}^{(0)}({\rm other}~0^-,0^+)~&=&~ -0.0015 \pm 0.0015~,
\eeqn
leading to the result
\beq
\label{eq_delta000}
         \Delta_\tau^{00}(J=0)~=~ 0.155 \pm 0.031~,
\eeq
where the $J=0$ contributions in the nonstrange part have been computed
to be $-0.00076$ for the $\pi$ pole (using the corresponding ALEPH 
branching ratio~\cite{alephbr_had}) and safely assumed to be negligible
in the other modes. Similarly, the $J=0$ contributions are obtained
for all ($k,l$) values and subtracted from the measured moments. 

\subsection{Theoretical parameters and uncertainties}

The recent ALEPH measurement of 
the strong coupling constant~\cite{alphas}, 
$\alpha_s(M_\tau^2) = 0.334\pm0.007_{\rm exp}\pm0.021_{\rm th}$~, 
is used. The theoretical error is dominated by uncertainties
due to the truncation of the perturbative series so that correlations
between the above value and the moments~(\ref{eq_dmoments})
are negligible, with the 
latter dominated by the uncertainties on the strange part. 

The dependence on the renormalization scheme (\MSbar)
used is already included in the theoretical error on \asm\ and thus 
not further added in order to avoid double counting.

The renormalization scale ($\mu$) dependence of the $D=2$
prediction has been studied in Ref.~\cite{pichprad}. When varying
$\mu$ away from $M_\tau$, additional logarithms enter the series and due to 
its truncation at finite order, a residual dependence on $\mu$ is
left. To estimate the associated uncertainty, $\mu$ is varied from 
1.3 to 3.6 GeV/$c^2$.

The uncertainty originating from the truncation of the perturbative
series for the dimension $D=2$ term is handled as follows:
the estimated error is taken as the full size
of the last term retained in the expansion, namely of order
$\alpha_s^3$ for the $J=1+0$ series.
Some redundancy is expected between the uncertainties estimated for
the renormalization scale dependence and the truncation of the
perturbative series. However, 
both errors are conservatively kept and added in quadrature.

For the dimension $D=4$ nonperturbative contribution
the products of quark condensates and quark masses are taken
from the PCAC relations with some correction for the strange 
quark~\cite{cdh}.
No errors are introduced for the higher dimensional operators
since they are fixed experimentally.
Finally, a small uncertainty is included from the error on $S_{\rm EW}$.

Table~\ref{tab_theoerr} gives the above theoretical uncertainties
on $m_s(M_\tau^2)$. 
The main error contribution stems from the $|V_{us}|$ 
uncertainty~\cite{pdg98}.
\begin{table}[t]
\centerline{
\begin{tabular}{|lcc|} 
\hline \hline
Error source &  Range &  $\Delta m_s(M_\tau^2)$~(MeV/$c^2$) \\
\hline\hline
$\alpha_s(M_\tau^2)$ & 
$0.334\pm0.022$   & 3  \\
truncation           &
                  & 11  \\ 
R-scale $\mu$     & $(1.3-3.6)$ GeV/$c^2$   
                  & 6 \\
$m_s\langle\bar{s}s\rangle$ 
                  & $-(1.63\pm0.29)\times10^{-3}~{\rm GeV}^4$
                  & 1 \\
$S_{\rm EW}$      & $1.0194\pm0.0040$ 
                  & 1  \\
$|V_{us}|$        & $0.2218\pm0.0016$ 
                  &22  \\
\hline
Total errors      &                  
                  &26 \\
\hline\hline
\end{tabular}
}
\caption[.]{\label{tab_theoerr}\it
              Theoretical uncertainties on
              $m_s(M_\tau^2)$. 
              The uncertainty from the truncation of the perturbative
              series for the $D=2$ mass term is estimated from adding
              or subtracting the value of the last retained term in
              the QCD expansion. The unequal positive and negative errors 
              on $m_s$ are averaged.
}
\end{table}

\subsection{Results}

The f\/it minimizes the $\chi^2$ of the dif\/ferences between measured 
and f\/itted quantities contracted with the inverse of the sum of the 
experimental and theoretical covariance matrices. Due to
the large correlations and the statistical limitation of the \sfs\
at the end of the $\tau$ phase space, higher moments $l>2$ do not
add significant information to the fit. 
The results for \msm\ and the nonperturbative 
contributions to $\Delta_\tau^{00(1+0)}$ for the `1+0' method are
\beqn
\label{res_ms}
   m_s(M_\tau^2)~&=&~(176^{ \,+37_{\rm exp} + 24_{\rm th}}
                   _{\,-48_{\rm exp} - 28_{\rm th}} 
                   \pm 8_{\rm fit} \pm 11_{J=0})~{\rm MeV}/c^2~, \\
\label{res_O6}
   \tilde{\delta}^{(6)}~&=&~0.039  \pm 0.016_{\rm exp}
                           \pm 0.014_{\rm th}\pm 0.004_{\rm fit}~,\\
\label{res_O8}
   \tilde{\delta}^{(8)}~&=&~-0.021 \pm 0.014_{\rm exp}
                           \pm 0.008_{\rm th}\pm 0.003_{\rm fit}~,
\eeqn 
with a $\chi^2/ndf$ of 0.2/2.

The errors are separated according to their experimental, 
theoretical, fit and spin separation origins. The fit error
stems from an intrinsic bias in the $\chi^2$ minimization,
due to the large correlations of the input observables, and includes
the total difference between the fit results with and without
correlations, while the results without correlations are given as
central values, as discussed in Ref.~\cite{dagostini}. 
\begin{table}[t]
\centerline{
\begin{tabular}{|lccc|} \hline\hline\rule{0pt}{13pt}
    & \msm\  & $\tilde{\delta}^{(6)}$ & $\tilde{\delta}^{(8)}$ \\ \hline
\msm\          &  1             &  0.74 & $-0.92$  \\
$\tilde{\delta}^{(6)}$ & --             &  1    & $-0.83$  \\
$\tilde{\delta}^{(8)}$ & --             &  --   & 1        \\
\hline\hline
\end{tabular}
}
\caption[.]{\label{tab_rescorr}\it
            Correlation matrix according to the `1+0' fit results.}
\end{table}
The correlation matrix corresponding to the fit results
is given in Table~\ref{tab_rescorr}.

The result (\ref{res_ms})
 \beq
    m_s(M_\tau^2) = (176^{\,+46}_{\,-57})~{\rm MeV}/c^2~
\eeq
can be evolved to the scale
of 1 GeV using the four-loop 
RGE $\gamma$-function~\cite{ritmass}, yielding
\beq
   m_s(1~{\rm GeV}^2) = (234^{\,+61}_{\,-76})~{\rm MeV}/c^2~.
\eeq

The fitted $m_s$ value corresponds to a contribution
$\delta^{(2)}_S=-(0.058\pm0.027)$ to $R^{(1+0)}_{\tau,S}$.
No operator of dimension $D=4$ has been fitted except the small
quartic strange mass corrections, for a total contribution 
$\delta^{(4)}_S=-(0.003\pm0.001)$, dominated by the strange quark 
condensate; the error given is almost entirely of
theoretical origin. The $D=6$ and $D=8$ strange contributions 
are $\delta^{(6)}_S=-0.038\pm0.022$ and
$\delta^{(8)}_S= 0.020\pm0.016$. They   
are fairly large compared to the nonstrange case, 
$\delta^{(6)}= 0.001\pm0.004$ and
$\delta^{(8)}=-0.001\pm0.001$~\cite{alphas}.

The above results are obtained from the $J=1+0$ piece which was shown
in Section 7.1.1 to have a satisfactory convergence in QCD. 
The inclusive method has a better experimental definition, but
suffers from the bad convergence properties of the $J=0$ part for
which a prescription must be specified. A reasonable rule to handle
an asymptotic series is to truncate it where the terms reach a minimum and
then assign as an uncertainty the full amount of the last term retained. 
With this prescription, the $J=0$ expansion is stopped after the $O(\alpha_s)$
term and, going through the same steps as before, one gets
\beq
\label{res_msinclu}
   m_s(M_\tau^2)~=~(149^{ \,+24_{\rm exp} + 21_{\rm th}}
                   _{\,-30_{\rm exp} - 25_{\rm th}} 
                   \pm 6_{\rm fit})~{\rm MeV}/c^2~, \\
\eeq 
where this time the theoretical uncertainty is dominated by the truncation
of the perturbative series. The fitted nonperturbative parts are
$\tilde{\delta}^{(6)}=0.031  \pm 0.015$
and $\tilde{\delta}^{(8)}=-0.018  \pm 0.011$.
The results (\ref{res_ms}) and
(\ref{res_msinclu}) are consistent within the uncorrelated part 
of their errors with a $\chi^2/ndf$ of 1.4/1.

It may be interesting to come back to the discarded $J=0$ part 
in order to get some information regarding the handling of its ill-behaved 
perturbative expansion, Eq.~(\ref{eq_cfoptci10}). Truncating the series at the
minimum provides a result consistent with (\ref{res_ms}), while keeping the
two more known terms destroys the consistency. This observation supports
the prescription used in the inclusive method. However, as stated above,
the safer determination with the `1+0' method is preferred and given as
the final result of this analysis.

\begin{figure}
\epsfxsize8cm
\centerline{\epsffile{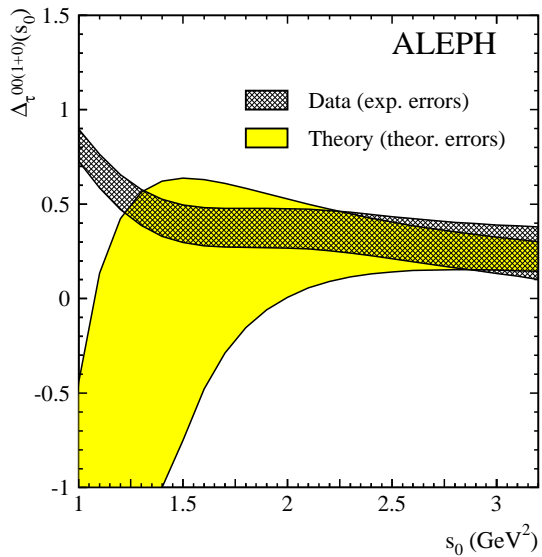}}
  \caption[.]{\label{fig_msrun}\it
              The observable $\Delta_\tau^{00(1+0)}(s_0)$
              as a function of the ``$\tau$-mass''-squared $s_0$. 
              The curve is  plotted
              as a one-standard deviation error band to emphasize
              its strong point-to-point correlations. Also shown
              is the theoretical prediction for the fit parameters 
              given in Eq.~(\ref{res_ms},\ref{res_O6},\ref{res_O8}).} 
\end{figure}

As shown for the nonstrange case in Ref.~\cite{alphas}, one
can simulate the physics of a hypothetical $\tau$ lepton with 
mass $\sqrt{s_0}$ smaller than $M_\tau$ by replacing $M_\tau^2$ everywhere
in Eq.~(\ref{eq_rtau}) by $s_0$. Under the 
assumption of quark-hadron duality, the evaluation of the observables as a
function of $s_0$ constitutes a test of the OPE approach adopted here,
since the energy dependence of the theoretical predictions is determined
once the parameters are fixed.
Figure~\ref{fig_msrun} shows the running observable 
$\Delta_\tau^{00(1+0)}(s_0)$ compared to the corresponding theoretical
predictions and the fitted parameters in 
Eq.~(\ref{res_ms},\ref{res_O6},\ref{res_O8}).
Despite the expected breakdown of the perturbative
approach at lower scales, it is noteworthy that 
the prediction agrees with data at such a low mass scale
within the theoretical uncertainties used in the analysis.

\subsection{Comparison with other determinations of $m_s$}

Other determinations of $m_s$ have been obtained by analyses of the
divergence of the vector and axial-vector current two-point function
correlators~\cite{gasser,jamin1,chetyrkin1,becchi,dominguez1,dominguez2,jamin2,kataev,colangelo,narison1,chetyrkin2}. 
The phenomenological information on the associated scalar and pseudoscalar
spectral functions is reconstructed from phase-shift resonance analyses which
are yet incomplete over the considered mass range 
and need to be supplemented by other assumed ingredients, 
in particular the description of the continuum, thus introducing
systematic effects.

Another approach~\cite{narison3} considers the difference between
isovector and hypercharge vector current correlators as related
to the $I=1$ and $I=0$ spectral functions accessible in \ee\
annihilation into hadrons at low energy. A recent reanalysis~\cite{maltman2}
points out the possibility of large corrections from isospin breaking,
leading to significant deviations for the extracted $m_s$ value.

Finally, lattice QCD calculations of $m_s$ 
are available~\cite{allton,gupta,eicker,gough}.
Their values show quite a large spread.

\begin{figure}[t]
\epsfxsize10cm
\centerline{\epsffile{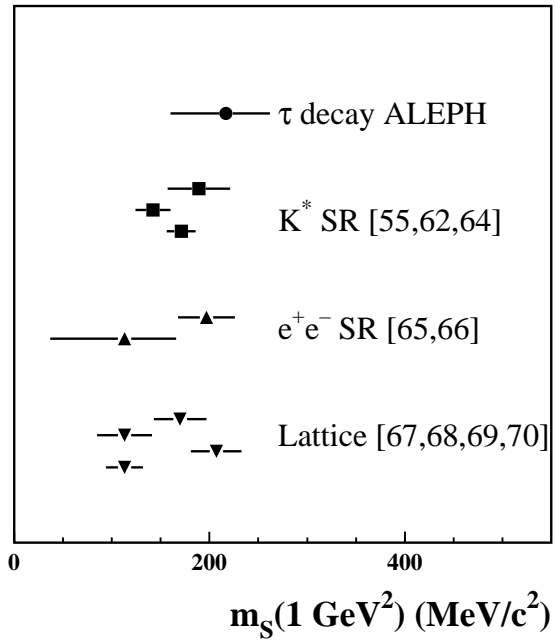}}
  \caption[.]{\label{compar}\it
              The ALEPH determination of 
$m_s(\mbox{1 GeV}^2)$ compared to the results 
              of other approaches. Details are given in the text 
(SR = sum rules). The references are listed in the order of
the results from top to bottom.
}
\end{figure}

The present determination of $m_s$ is directly compared with the other
derivations in Fig.~\ref{compar} where all values are given 
at a mass scale of 1 GeV. Qualitative agreement is observed, the value
from the present analysis being rather on the high side of the
range of previous determinations. The precision of the result is 
unfortunately still limited. As an example, 
the Standard Model prediction for the CP-violation 
parameter $\epsilon '/\epsilon$ is strongly dependent on the
value of $m_s$ and the present determination translates into a range of
$\epsilon '/\epsilon$ values spanning a factor of five~\cite{buras},
but favouring ``low'' values for this ratio. 

\section{Conclusion}

All ALEPH measurements on $\tau$ decays with kaons are summarized
to provide an overall review and a comprehensive study of these decays.

Comparing the decays $\tau^-\to K^-\nu_\tau$ and
$K^-\to \mu^-\nu_\tau$ shows agreement with $\mu-\tau$ universality
within $2\%$ uncertainty. 
 
The investigation of mass spectra 
shows that the $(\overline{K}\pi)^-$ decay mode is dominated by $K^*(892)^-$.
A $K^{*}(1410)$ contribution is extracted from 
a fit to the $\overline{K}\pi$ mass spectrum through its interference with 
the dominant $K^*(892)$. 

The $(\overline{K}\pi\pi)^-$ decay modes are observed to
proceed through the $K_1(1270)$, $K_1(1400)$ and $K^{*}(1410)$ resonances.
The relative contributions are extracted by fitting the $\pi\pi$ and
$\overline{K}\pi\pi$ mass spectra, separating the vector 
and axial-vector contributions. The vector current contribution to
the $\overline{K}\pi\pi$ modes is found to be
$(23^{\,+22}_{\,-17})\%$.

According to CVC, the vector spectral function for the $(K\overline{K}\pi)^-$
modes is connected to low energy $e^+e^-$ annihilation, allowing one to 
conclude that $(94^{\,+6}_{\,-8})\%$ of this decay mode proceeds  
through the axial-vector current. Because the latter is  
dominated by the $a_1$ resonance, a branching ratio
$B(a_1 \to K^*K)= (2.6 \pm 0.3)\%$ is obtained.

Tests of isospin symmetry for 
the $(\overline{K}\pi)^-$, $(\overline{K}\pi\pi)^-$ and 
$(K\overline{K}\pi)^-$ modes are performed. A good consistency 
with the expectation from isospin invariance is observed,
providing useful information on the decay dynamics in these final states.  

The branching ratio for $\tau$ decay into all strange final states
is determined to be $B(\tau^-\to X^-(S=-1)\nu_\tau)=(28.7\pm1.2)\times 10^{-3}$
and the total strange spectral function is derived from
the corresponding mass spectra.

Using the total rates, as well as moments of the respective spectral 
functions, through a combination of strange and nonstrange parts which
cancel the dominant massless perturbative QCD contribution, a global fit
is performed providing a determination of the strange nonperturbative 
contributions and of the $s$ quark mass. The value obtained is 
$m_s(M^2_\tau)=(176^{\,+46}_{\,-57})~\mbox{MeV}/c^2$, which is 
evolved to 1 GeV to yield 
$m_s(1~{\rm GeV}^2) = (234^{\,+61}_{\,-76})~{\rm MeV}/c^2$, in
agreement with other determinations.

\section*{Acknowledgements}

We are indebted to K.G. Chetyrkin,  
J.H. K\"{u}hn, K. Maltman, A. Pich and J. Prades for useful discussions. 
We also wish to thank our colleagues in the CERN accelerator divisions for the
successful operation of the LEP storage ring. We thank the engineers
and technicians in all our institutions for their support in
constructing and operating ALEPH. Those of us from nonmember states
thank CERN for its hospitality.

\end{document}